\documentclass[aps,prd,notitlepage,superscriptaddress,longbibliography,nofootinbib]{revtex4-1}

\usepackage{dcolumn}    % 小数点揃え
\usepackage{bm}         % 太字ベクトル
\usepackage{hyperref}   % ハイパーリンク
\usepackage{bmpsize}
\usepackage[english]{babel}
\usepackage[skip=1pt, justification=raggedright]{caption}
\usepackage{graphicx}
\usepackage{enumitem}
\vspace{-10pt}
\usepackage{amsmath}
\hypersetup{
    colorlinks=true,
    citecolor=blue,
    linkcolor=blue,
    urlcolor=blue
}

\begin{document}

\title{Cryogenic sub-Hz cROss torsion bar detector with quantum NOn-demolition Speed meter (CHRONOS) for gravitational wave detection}

\author{Yuki Inoue}
\thanks{Corresponding author: iyuki@ncu.edu.tw}
\affiliation{Department of Physics,National Central University, Taoyuan, Taiwan}
\affiliation{Center for High Energy and High Field (CHiP), National Central University,  Taoyuan, Taiwan}
\affiliation{Institute of Physics, Academia Sinica, Taipei, Taiwan}
\affiliation{Institute of Particle and Nuclear Studies, High Energy Acceleration Research Organization (KEK), Tsukuba, Japan}

\author{Daiki Tanabe}
\affiliation{Institute of Physics, Academia Sinica, Taipei, Taiwan}
\affiliation{Center for High Energy and High Field (CHiP), National Central University,  Taoyuan, Taiwan}
\affiliation{Institute of Particle and Nuclear Studies, High Energy Acceleration Research Organization (KEK), Tsukuba, Japan}

\author{M.Afif Ismail}
\affiliation{Department of Physics,National Central University, Taoyuan, Taiwan}
\affiliation{Center for High Energy and High Field (CHiP), National Central University,  Taoyuan, Taiwan}
\affiliation{Institute of Physics, Academia Sinica, Taipei, Taiwan}

\author{Vivek Kumar}
\affiliation{Department of Physics,National Central University, Taoyuan, Taiwan}
\affiliation{Center for High Energy and High Field (CHiP), National Central University,  Taoyuan, Taiwan}

\author{Mario Juvenal S Onglao III}
\affiliation{National Institute of Physics, University of the Philippines - Diliman, Quezon City 1101, Philippines}
\affiliation{Center for High Energy and High Field (CHiP), National Central University, Taoyuan, Taiwan}

\author{Ta-Chun Yu}
\affiliation{Department of Physics,National Central University, Taoyuan, Taiwan}
\affiliation{Center for High Energy and High Field (CHiP), National Central University,  Taoyuan, Taiwan}

\date{\today}

\begin{abstract}

We present the optical design and sensitivity modeling of the 2.5 m Cryogenic sub-Hz cROss torsion-bar detector with quantum NOn-demolition Speed meter (CHRONOS), a triangular Sagnac speed-meter interferometer incorporating power- and signal-recycling techniques.
Using ABCD-matrix analysis and \textsc{Finesse3} simulations, we show that stable eigenmodes are obtained with optimized mirror curvatures and focal placements, achieving mode-matching efficiencies above 99.5\%.
The configuration reaches a quantum-noise-limited strain sensitivity of
$h \simeq 3\times10^{-18}\,\mathrm{Hz^{-1/2}}$ at 1 Hz,
with a ring-cavity finesse $\mathcal{F}\simeq3.1\times10^{4}$ and a round-trip Gouy phase $\psi\approx153^{\circ}$.
The power-recycling cavity detuning ($\phi_p=5^{\circ}$) dominates the low-frequency quantum noise, while the signal-recycling cavity detuning ($\phi_s=0^{\circ}$) mainly introduces a uniform quadrature rotation.
The optimal homodyne angle ($\zeta\simeq46^{\circ}$) balances shot noise and radiation-pressure effects to give the best sensitivity near 1 Hz.
Assuming an end-mirror reflectivity $R_{\mathrm{ETM}}=99.9999\%$ under cryogenic operation at 10 K, CHRONOS can achieve quantum-noise-limited performance on a laboratory scale and serve as a testbed for future long-baseline cryogenic interferometers probing sub-hertz gravitational waves.
The projected science reach includes the detection of intermediate-mass
black-hole binaries with total masses of $10^{2}$--$10^{5}\,M_{\odot}$,
reaching a maximum detection distance of $\sim271\,{\rm Mpc}$ for binaries
with total masses around $9\times10^{3}\,M_{\odot}$ (SNR = 1). The detector
also achieves a best power-integrated sensitivity to stochastic
gravitational-wave backgrounds of
$\Omega_{\rm GW}\simeq4.7\times10^{-4}$ at $2.15\,{\rm Hz}$ after 10 years of
observation, probes Yukawa-type deviations from Newtonian gravity down to
$|\alpha|\sim2\times10^{-5}$, and can detect prompt gravity-gradient signals
from nearby M5.2 earthquakes out to distances of approximately $90\,{\rm km}$.
\end{abstract}
\maketitle

\section{Introduction}
Gravitational-wave astronomy has advanced rapidly through the success of ground-based interferometers such as LIGO, Virgo, and KAGRA~\cite{Abbott2016,GWTC3,KAGRA2020}. Since the first direct detection of gravitational waves from GW150914, the field has evolved from demonstrating the existence of gravitational waves to establishing gravitational-wave astronomy as a precision observational science. Over the past decade, the growing catalog of compact-binary mergers has highlighted the importance of exploring previously inaccessible frequency bands. Meanwhile, the observational window is being extended into the millihertz regime by the upcoming space-based Laser Interferometer Space Antenna (LISA)~\cite{LISA2017}. However, a sensitivity gap remains in the sub-hertz band. Ground-based detectors are fundamentally limited at low frequencies by seismic and environmental noise, while space missions cannot adequately cover this regime due to practical constraints on arm length. As gravitational-wave astronomy enters its second decade, extending observational capability toward lower frequencies has become one of the major challenges for the next generation of detectors.

To access this unexplored frequency range, several terrestrial detector concepts have been proposed and developed, including the torsion-bar antenna (TOBA)~\cite{Ando2010} and the Torsion Pendulum Dual Oscillator (TorPeDO)~\cite{McManus2017TorPeDO}, both of which have demonstrated proof-of-principle operation of sub-hertz gravitational-wave detectors. At the same time, substantial advances have been achieved in cryogenic engineering, ultra-low-loss optical materials, precision interferometry, and quantum-noise reduction techniques. CHRONOS builds upon these developments by combining cryogenic torsion-bar test masses with a quantum non-demolition (QND) Sagnac speed-meter interferometer, aiming to bridge the sensitivity gap between terrestrial and space-based observatories in the sub-hertz band~\cite{5zyh-lg9b}.

Opening the sub-hertz band would unlock new opportunities for gravitational-wave astronomy. 
In particular, the merger signals of intermediate-mass black hole binaries exhibit characteristic frequency evolution in this regime. 
For the Cryogenic sub-Hz cROss torsion bar detector with quantum NOn-demolition Speed meter (CHRONOS) sensitivity scale~\cite{inoue2025chronos, White_paper_CHRONOS, SPP-2026-3A-06}, a facility could detect mergers with total mass $M \sim 10^4 M_\odot$, 
enabling multi-band observations that bridge LISA and future ground-based detectors. 
Such observations would provide direct evidence for intermediate-mass black holes and valuable insights into the formation pathways of supermassive black holes~\cite{AmaroSeoane2017}. 
Moreover, recent O4 observations have begun to reveal the black-hole mass distribution, 
with several candidate intermediate-mass black hole events already reported~\cite{GWTC3,GWTC4pop}. 
Measurements in the high-mass regime are particularly important for understanding black-hole formation scenarios, 
and CHRONOS will offer unique access to this parameter space.  

In addition, CHRONOS provides unique sensitivity to the stochastic gravitational-wave background $\Omega_{\rm GW}$ of cosmological origin. 
With 10 years of integrated observation, the detector is expected to reach a level of $\Omega_{\rm GW} \sim 5 \times 10^{-4}$ at 2~Hz. 
This would directly probe early-universe phenomena such as first-order phase transitions and cosmic strings, 
thus offering rare opportunities to test physics beyond the standard cosmological model~\cite{Caprini2016,Ng2021,inoue2025chronos}.  

The broader scientific prospects and overall detector concept are discussed in detail in the companion CHRONOS design paper. 
In this study, we report on the first stage of development: the optical design and sensitivity evaluation. 
The CHRONOS adopts a triangular Sagnac configuration with output recycling to enhance sensitivity. 
We investigate suitable sideband signals for controlling multiple cavities and present design parameters for realistic optical configurations such as dual-recycling cavities. 
To achieve stable locking of the Sagnac cavity, we compute the mirror-curvature conditions required to reach the desired mode-matching efficiency (99.5\%), 
and confirm consistency across the full optical design. 
Finally, we calculate sensitivity curves based on the concrete optical layout and discuss the fundamental sensitivity limits and science of CHRONOS.

\section{Principle}
\begin{figure}[t]
  \centering
  \includegraphics[width=0.7\linewidth]{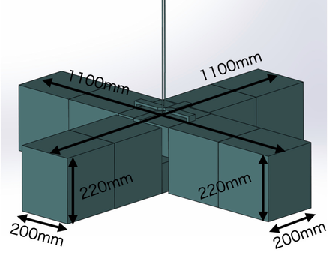}
  \caption{
    Schematic of the bar-shaped test masses.
    Each bar has dimensions of $1100\,\mathrm{mm}$ in length, $200\,\mathrm{mm}$ in width, and $220\,\mathrm{mm}$ in height.
    Two orthogonal bars (X and Y) are suspended at the center to realize differential torsional readout.
  }
  \label{fig:test_mass}
\end{figure}

The readout in this study follows the TOBA framework~\cite{Ando2010}, 
which is based on the torsional degree of freedom around the $z$ axis.
The angular displacement $\theta$ of a bar-shaped test mass with moment of inertia $I$, 
torsional spring constant $\kappa$, and loss coefficient $\gamma$ obeys
\begin{equation}
    I\,\ddot{\theta}(t) + \gamma\,\dot{\theta}(t) + \kappa\,\theta(t)
    = N_{\mathrm{gw}}(t),
    \label{eq:eom_toba_time}
\end{equation}
where the torque from the gravitational-wave tidal acceleration 
$a_i = \tfrac{1}{2}\,\ddot{h}_{ij}\,x^j$ is given by~\cite{Saulson1994,Forward1978}
\begin{equation}
    N_{\mathrm{gw}}(t) = \frac{1}{2}\,\ddot{h}_{ij}(t)\,\epsilon_{zki}\,Q_{kj},
\end{equation}
with
\begin{equation}
    Q_{ij} \equiv \int_V \rho\!\left(x_i x_j - \tfrac{1}{3} r^2 \delta_{ij}\right)\,dV,
\end{equation}
where the dimension of the torsion bar is described in Fig.~\ref{fig:test_mass}. 
In the frequency domain, this becomes
\begin{align}
    \theta(\Omega)
    &= \chi_\theta(\Omega)\!\left[
       -\frac{I_{\mathrm{eff}}}{2}\Omega^2\bigl(h_{+}F_{+} + h_{\times}F_{\times}\bigr)
     \right], \\
    \chi_\theta(\Omega) &\equiv \frac{1}{\kappa - I\Omega^2 + i\gamma\Omega},
    \label{eq:TF}
\end{align}
where the torsional resonance $\Omega_t = \sqrt{\kappa/I}=2\pi \times 0.004$ is taken to be well below the observational band (0.1--10 Hz).
In the regime $\Omega \gg \Omega_t$, the susceptibility reduces to $\chi_\theta \simeq (-I\Omega^2)^{-1}$, yielding
\begin{equation}
    \theta_i(\Omega) \simeq \frac{\eta_g}{2}\bigl(h_{+}F_{+} + h_{\times}F_{\times}\bigr),
    \label{eq:theta_flat}
\end{equation}
where $i = \{X,Y\}$.
The total and effective moments of inertia are
\begin{equation}
    I = \int \rho(x^2 + y^2)\,dV, \qquad
    I_{\mathrm{eff}} = \int \rho(x^2 - y^2)\,dV,
\end{equation}
such that the geometrical coupling factor is defined as $ \eta_g \equiv \frac{I_{\mathrm{eff}}}{I}$.
For the present CHRONOS test-mass geometry, we obtain
$\eta_g = 0.936$.
In the long-wavelength limit with $z$-axis incidence, the antenna patterns are~\cite{Forward1978}
$ F_{+} = \sin 2\alpha$, $F_{\times} = -\cos 2\alpha .$
By combining two orthogonal bars (X and Y) into a differential readout,
the antenna responses acquire opposite signs because the two bars are rotated by $90^\circ$ with respect to each other.
Consequently, the differential channel provides an effective factor-of-two enhancement of the gravitational-wave signal,
\begin{equation}
    \theta(\Omega) = \theta_X - \theta_Y
    = \eta_g \bigl(h_{+}F_{+} + h_{\times}F_{\times}\bigr),
    \label{eq:theta_diff}
\end{equation}
which suppresses common-mode disturbances, including residual translational motion and a fraction of laser intensity noise,
while retaining sensitivity to linear combinations of the two polarizations~\cite{Ando2010}.

While the differential torsional readout provides direct sensitivity to
gravitational waves, a conventional angle measurement is fundamentally
limited by quantum back-action arising from radiation-pressure fluctuations.
In a position-meter configuration, the observable is the angular displacement
$\theta$, whose future evolution is perturbed by the measurement process.
This leads to the well-known trade-off between shot noise and
radiation-pressure noise, ultimately resulting in the Standard Quantum Limit.

To overcome this limitation, CHRONOS adopts a Sagnac speed-meter
interferometer. As shown in Fig.~\ref{fig:layout},
counter-propagating beams circulate in the triangular cavities and are recombined at Beam splitter.
Because the two beams probe the test mass at different times separated by
the optical storage time $\tau$, the output signal is proportional to

\begin{equation}
s(t)
\propto
\frac{\theta(t+\tau)-\theta(t)}{\tau}. \label{eq:SSM}
\end{equation} 

For frequencies below the inverse delay time,
the finite difference approaches the time derivative of the angular
displacement, $s(t) \propto \dot{\theta}(t).$

In the observational band of CHRONOS, the torsional resonance frequency
$\Omega_t$ is designed to be far below the measurement band
($\Omega \gg \Omega_t$).
The restoring torque therefore becomes negligible and the torsion bar behaves
approximately as a free rotor.
The Hamiltonian can then be written as

\begin{equation}
\hat H
\simeq
\frac{\hat L^2}{2I},
\end{equation}

where $\hat L$ denotes the angular momentum operator.
Since $[\hat L,\hat H]=0,$

the angular momentum is conserved,

\begin{equation}
\frac{d\hat L}{dt}
=
\frac{1}{i\hbar}
[\hat L,\hat H]
=
0.
\end{equation}

Consequently, $[\hat L(t_1),\hat L(t_2)] = 0,$
which satisfies the Braginsky criterion for a QND
observable.

In this limit, the measured quantity is directly related to the angular
momentum through $L \simeq I\dot{\theta},$
so that the Sagnac speed meter provides access to a QND observable rather
than the displacement itself.
Consequently, the measurement back-action can be substantially reduced
compared with a conventional position-meter configuration.

The angular momentum and angular displacement satisfy the canonical
commutation relation $[\hat L,\hat \theta] \simeq i\hbar,$

and therefore obey Ozawa's generalized error--disturbance relation~\cite{Ozawa2003, Erhart2012, Rozema2012}

\begin{equation}
\epsilon(L)\eta(\theta)
+
\sigma(L)\eta(\theta)
+
\epsilon(L)\sigma(\theta)
\ge
\frac{\hbar}{2}.
\end{equation}

By enabling a macroscopic QND measurement of angular momentum with
$\mathcal{O}(100\,\mathrm{kg})$ test masses, CHRONOS provides a unique
experimental platform for exploring quantum measurement theory in a regime
far beyond that accessible in microscopic systems.

The detailed quantum-noise model of the Sagnac speed meter, including the
suppression of radiation-pressure noise and the resulting sensitivity
improvement, is presented in Sec.~\ref{sec:sensitivity}.

%%%%%%%%%%%%%%%%%%%%%
\section{Optical layout}
The previous section introduced the operating principle of CHRONOS,
namely a torsion-bar speed-meter interferometer that measures
the angular velocity of macroscopic test masses using a Sagnac topology.
To realize this concept experimentally,
a practical optical configuration is required that simultaneously provides
stable cavity eigenmodes, efficient optical power buildup,
and quantum-noise-limited readout.
In this section, we present the baseline optical design adopted for the
first CHRONOS prototype.

An overview of the optical layout is shown in
Fig.~\ref{fig:layout}.
The optical system can be divided into five major subsystems:
(i) the triangular ring cavities that sense the torsional motion,
(ii) the Sagnac interferometer that realizes the speed-meter response,
(iii) the power-recycling cavity for optical power enhancement,
(iv) the signal-recycling cavity for quantum-noise optimization,
and (v) the input and output optical systems used for beam preparation
and signal readout.
These subsystems are highlighted using different colors in
Fig.~\ref{fig:layout}.

The experimental apparatus is based on a cross-torsion-bar configuration,
with identical optical systems placed in the X and Y arms.
For convenience, the following description focuses on the Y arm,
although the X arm has an identical optical configuration.

After pre-stabilization, the input laser beam first passes through
a pre-stabilized laser system and an input mode cleaner,
which stabilize and shape the spatial and frequency profiles
\cite{aasi2015advanced,abbott2021ligo}.
The beam is then split at the input beam splitter (IBS)
and directed into the Sagnac interferometers of the X and Y arms.
Within each arm, the light propagates through triangular ring cavities,
and the returning beams are recombined at the output beam splitter (OBS),
where the differential phase signal is extracted by balanced homodyne readout method.

\subsection{Triangular ring cavity}
The central sensing element of CHRONOS is a set of triangular ring cavities,
highlighted by the blue labels in Fig.~\ref{fig:layout}.
Each cavity is formed by an input test mass ($ITM_{ij}$),
a middle test mass ($MTM_{ij}$),
and an end test mass ($ETM_{ij}$),
and is installed at the ends of the cross-shaped torsion bars, where $i = \{L, R\}$ and $j =\{ X, Y \}$.
These cavities provide the optical interaction region in which
the rotational motion of the torsion bars is converted into an optical phase signal.

To support stable cavity eigenmodes,
the $ETM_{ij}$ and cavity mirrors ($CM_j$) are assigned finite radii of curvature.
These curved mirrors define the cavity stability and establish
an internal beam waist within the ring cavity.
The resulting optical mode serves as the fundamental spatial mode
for all subsequent mode-matching calculations.

In a simplified description,
the optical path of each ring cavity is defined by six mirrors,
including the $ITM_{Lj}$ and $ITM_{Rj}$,
forming a closed traveling-wave resonator.
This geometry enables the realization of a Sagnac speed-meter interferometer,
which will be described in the next subsection.

\subsection{Sagnac speed-meter interferometer}

The triangular ring cavities described above are combined to form
a Sagnac speed-meter interferometer,
highlighted by the green labels in Fig.~\ref{fig:layout}.
The Sagnac topology is a key feature of CHRONOS,
as it enables direct measurement of test-mass velocity
rather than displacement.
This property provides a natural suppression of quantum
radiation-pressure noise at low frequencies,
making it particularly attractive for sub-hertz gravitational-wave detectors
\cite{chen2003,danilishin2012}.

After passing through the beam splitter ($BS_j$),
the optical field is divided into clockwise(CW) and counterclockwise(CCW)
propagating beams.
These two beams traverse the ring cavities in opposite directions
and subsequently recombine at the output beam splitter.
Because each beam samples the motion of the test masses at different times,
the resulting interference signal is proportional to the temporal variation
of the mirror position.
At frequencies below the cavity pole,
this response approaches a measurement of test-mass velocity,
which is the characteristic feature of a speed-meter interferometer as described in Eq.\eqref{eq:SSM}.

As discussed in the previous section,
the speed-meter response suppresses the back-action noise that limits
conventional position meters.
This feature is particularly important for CHRONOS,
where quantum radiation-pressure noise becomes significant
in the target observation band between $0.1~\mathrm{Hz}$ and $10~\mathrm{Hz}$.

In each arm, two ring cavities are installed on opposite sides
of the torsion bar.
The differential phase accumulated in these cavities
encodes the rotational motion of the bar,
while common-mode fluctuations are largely rejected.
The signals from the X and Y interferometers are subsequently combined
to form the final gravitational-wave readout.

\subsection{Power-recycling cavity}
Although the Sagnac topology provides a favorable quantum-noise response,
its sensitivity is ultimately limited by the available circulating optical power.
To increase the intracavity power without requiring excessively high laser power,
CHRONOS employs a power-recycling cavity,
highlighted by the pink labels in Fig.~\ref{fig:layout}.

The power-recycling cavity is formed by the power-recycling mirrors
$PRM1_{j}$--$PRM4_{j}$ together with the central beam splitter.
Light returning toward the input port is resonantly recycled,
thereby increasing the circulating power inside the interferometer.
This enhancement improves the shot-noise-limited sensitivity
while maintaining a realistic input laser power.

Among the power-recycling mirrors,
$PRM2_{j}$ is assigned a finite radius of curvature,
whereas the remaining mirrors are planar.
The curved $PRM2_{j}$ mirror acts as a mode-matching element
between the external input beam and the eigenmode of the Sagnac ring cavities.
By adjusting the curvature of $PRM2_{j}$,
the position and size of the external beam waist can be controlled,
allowing efficient coupling into the interferometer.

This degree of freedom plays an important role in the optical design.
In the present study,
the $PRM2_{j}$ curvature is optimized to maximize the overlap
between the injected Gaussian beam and the cavity eigenmode.
The resulting mode-matching performance is evaluated quantitatively
in Sec.~\ref{sec:mode_matching}.

The power-recycling cavity increases the circulating optical power and gives
additional control of the interferometer response.
This subsystem is described in Sec.~\ref{sec:sensitivity}.

%%%%%%%%%%%%%%
\subsection{Signal-recycling cavity}
While the power-recycling cavity increases the circulating optical power,
the frequency response of the interferometer is further controlled
by a signal-recycling cavity,
highlighted by the purple labels in Fig.~\ref{fig:layout}.
Signal recycling is widely used in modern gravitational-wave detectors
to shape the quantum-noise spectrum and optimize the detector sensitivity
for a target frequency band~\cite{mizuno1993,kimble2001}.

The  signal-recycling cavity is formed by the signal-recycling mirrors
$SRM1_{j}$--$SRM3_{j}$ together with the central beam splitter.
Light carrying the gravitational-wave signal is partially reflected
back into the interferometer,
allowing the optical response to be modified through resonant enhancement
of selected signal sidebands.

For CHRONOS, signal recycling is particularly important because
the target observation band lies in the sub-hertz regime,
where both quantum noise and cavity dynamics strongly influence
the detector response.
By adjusting the microscopic detuning of the  signal-recycling cavity,
the balance between low-frequency sensitivity
and high-frequency response can be optimized.

The signal-recycling cavity therefore provides an additional degree of freedom
for tailoring the quantum-noise spectrum.
In combination with the Sagnac speed-meter topology,
it enables further suppression of quantum noise
within the frequency range relevant for CHRONOS observations.
The quantitative impact of signal recycling on the sensitivity curve
is discussed in Sec.~\ref{sec:impact_SRCD}.

Having defined the main interferometer cavities,
the remaining optical subsystems are responsible for preparing the input beam
and extracting the output signal with high fidelity.
These systems are described in the following subsection.

\subsection{Input and output optics}

\begin{figure}[]
  \centering
  \includegraphics[width=1.0\textwidth]{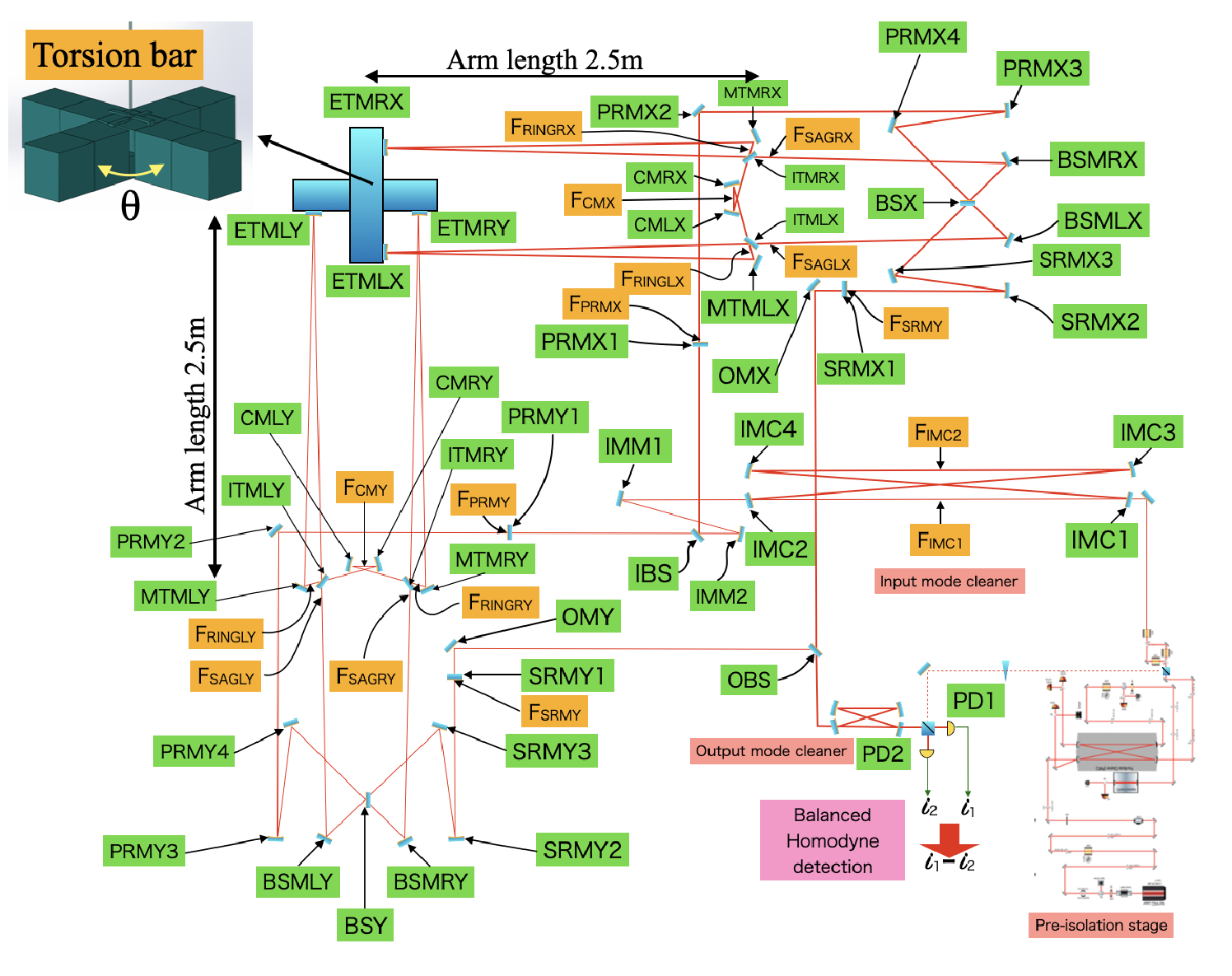}
\caption{
Optical layout of the CHRONOS speed-meter interferometer.
The optical system consists of five major subsystems:
the triangular ring cavities (blue),
the Sagnac interferometers (green),
the power-recycling cavity (pink),
the signal-recycling cavity (purple),
and the input/output optical systems (yellow and gray).
The input laser is first filtered by the input mode cleaner (IMC) and injected into the interferometer through the power-recycling cavity.
Inside each arm, light circulates in triangular ring cavities attached to the torsion-bar test masses, forming a Sagnac speed-meter interferometer that measures angular velocity rather than displacement.
The signal-recycling cavity modifies the interferometer response and enables optimization of the quantum-noise spectrum in the sub-hertz band.
At the output port, an output mode cleaner (OMC) suppresses higher-order spatial modes before the signal is read out using balanced-homodyne detection (BHD).
For clarity, only the Y-arm optical path is highlighted in detail, while the X arm has an identical configuration.
The red markers indicate the reference port and focal points
used for mode-matching design.
The internal focal points define the eigenmodes of the triangular
ring cavities, while the external reference port provides the
target Gaussian beam for injection and balanced-homodyne readout.
}
  \label{fig:layout}
\end{figure}

The remaining optical subsystems are responsible for preparing the input beam
and extracting the interferometric signal with high fidelity.
These components are highlighted by the yellow and gray labels in
Fig.~\ref{fig:layout}.

Before entering the interferometer,
the laser beam passes through a pre-stabilization system
and an input mode cleaner, $IMC_1$ -- $IMC_4$,
highlighted by the yellow labels in Fig.~\ref{fig:layout}.
The input mode cleaner, acts as a spatial and frequency filter,
suppressing beam jitter, higher-order spatial modes,
and residual laser-frequency noise
\cite{aasi2015advanced,abbott2021ligo}.
This filtering establishes a stable fundamental Gaussian mode
for injection into the power-recycling cavity.

At the output port,
an output mode cleaner is installed before the photodetectors.
The output mode cleaner removes higher-order spatial modes
and residual control sidebands that do not contribute
to the gravitational-wave signal.
For the present design,
a suppression level of approximately
$-40~\mathrm{dB}$ is assumed
\cite{aasi2015advanced},
ensuring that the detected optical field
is dominated by the fundamental signal mode.

After spatial filtering by the output mode cleaner,
the interferometer output is measured using
balanced-homodyne detection,
highlighted by the gray labels in Fig.~\ref{fig:layout}.
In this scheme,
the interferometer output is combined with a local oscillator beam,
and the photocurrents from two photodiodes are subtracted.
This approach enables measurement of an arbitrary optical quadrature
and provides improved quantum-noise performance
compared with conventional DC readout
\cite{fricke2012,vahlbruch2010}.

The homodyne angle serves as an important optimization parameter
for balancing shot noise and radiation-pressure noise.
Its impact on the overall detector sensitivity
will be investigated in Sec.~\ref{subsec:homodyne_angle}.

Several focal points are indicated in Fig.~\ref{fig:layout}
to visualize the Gaussian beam propagation through the system as highlighted by the orange labels in Fig.~\ref{fig:layout}.
The internal focal points correspond to beam waists formed by
the curved mirrors within the triangular ring cavities.
These waists define the cavity eigenmodes and serve as the reference
for the optical design.

Reference ports equipped with photodetectors are defined outside the interferometer, as highlighted by the red labels in Fig.~\ref{fig:layout}. These ports are used to demodulate the modulation sidebands injected via the input optics. The resulting error signals are indispensable for maintaining the interferometer lock, and the optimization of the modulation frequencies is described in Sec.~\ref{sec:cavity_control}.

\subsection{Summary of optical configuration}

In this paper, we focus on the interferometer,
which serves as the first experimental realization of the optical architecture described above.
All mirrors, beam splitters, and mode cleaners are fully specified in Appendix~\ref{app:optics}.
Table~\ref{tab:alloptics} summarizes the coordinates, incidence angles, radii of curvature, and reflectivities, allowing the optical layout shown here to be exactly reproduced.
These parameters provide the foundation for the beam-propagation simulations and sensitivity analysis presented in later sections.

Designing the optical layout alone is not sufficient to ensure practical interferometer operation.
To realize the proposed configuration, two additional requirements must be satisfied.
First, all optical cavities must be stably controlled and maintained on resonance, requiring appropriate cavity sensing and control schemes.
Second, efficient mode matching must be achieved between the external input beam, the recycling cavities, and the Sagnac ring cavities in order to maximize the circulating power and signal extraction efficiency.

The following two sections address these requirements.
Section~\ref{sec:cavity_control} presents the cavity control strategy and evaluates the sensing signals required for stable operation,
while Sec.~\ref{sec:mode_matching} investigates the Gaussian-beam propagation and mode-matching performance of the optical system.

%%%%%%%%%%%%%%
% Cavity control
%%%%%%%%%%%%%%

\section{Cavity control}
\label{sec:cavity_control}

Before discussing the cavity-control architecture of the interferometer,
it is useful to clarify the design philosophy that guided the selection of the optical topology,
cavity lengths, recycling configuration, and control sidebands adopted in this study.
The ultimate goal of CHRONOS is the realization of a large-scale sub-hertz gravitational-wave detector.
The interferometer serves as a system designed to validate the key optical
and control concepts required for such a detector while remaining experimentally feasible
within a laboratory environment.
Achieving this performance, however, requires stable operation of multiple coupled optical cavities,
precise mode matching, and robust control of the interferometer operating point.

The primary objective of the cavity-control system is therefore to maintain resonance conditions
for the Sagnac cavity and the recycling cavities while simultaneously providing stable references
for balanced-homodyne readout and quantum-noise measurements.

Each control degree of freedom, sensing port, and modulation sideband frequency
used in the present design is summarized in Table~\ref{tab:control_ports}.

\subsection{Sideband frequencies}

The selection of the radio-frequency (RF) sidebands is governed by the resonance conditions required for independent sensing and control of the power-recycling cavity, signal-recycling cavity, and triangular Sagnac cavity.
The sideband structure of CHRONOS is designed to reproduce the same resonant and anti-resonant relationships employed in large-scale gravitational-wave detectors.

The starting point of the design is the round-trip length of the triangular Sagnac ring cavity,
$
L_{\mathrm{ring}} = 5.08~\mathrm{m},
$
which corresponds to a free spectral range of
$
\mathrm{FSR}_{\mathrm{ring}}
=
\frac{c}{L_{\mathrm{ring}}}
\simeq
59.0~\mathrm{MHz}.
$

This frequency serves as the fundamental reference for the entire sideband scheme.
The lengths of the recycling cavities are then chosen such that specific sidebands experience different resonance conditions in the power-recycling cavity and signal-recycling cavity, enabling independent sensing of the corresponding longitudinal degrees of freedom.

The signal-recycling cavity is designed to have an effective free spectral range equal to one half of the ring free spectral range,
$
L_{\mathrm{src}}
\simeq
5.08~\mathrm{m},
$ $
\mathrm{FSR}_{\mathrm{src}}
\simeq \frac{c}{2L_{src}}
= 29.5~\mathrm{MHz},
$
while the power-recycling cavity is set to
$
L_{\mathrm{prc}}
\simeq
7.62~\mathrm{m}$, 
 $
\mathrm{FSR}_{\mathrm{prc}}
\simeq \frac{c}{2L_{prc}}
19.7~\mathrm{MHz}.
$

These lengths reproduce the resonance hierarchy originally developed for Advanced LIGO, in which the recycling cavities provide distinct optical responses to different RF sidebands.
Although the present interferometer is only 2.5~m in scale, the resulting optical phase relationships are equivalent to those of kilometer-scale detectors, allowing realistic validation of the cavity-control architecture.

Based on these resonance conditions, two RF sidebands are selected for the main interferometer control,
$
f_{1}=29.5~\mathrm{MHz},$
$f_{2}=59.0~\mathrm{MHz}.$

The frequency \(f_{1}\) corresponds to the signal-recycling cavity free spectral range, whereas \(f_{2}\) coincides with the ring-cavity free spectral range.
{\color{black}
As a result, the two sidebands experience different resonance conditions within the coupled cavity system.
In particular, \(f_{2}\) is resonant in the power-recycling cavity, while remaining anti-resonant in the signal-recycling cavity, providing strong discrimination between the two recycling-cavity degrees of freedom.
}
This separation enables efficient sensing of the longitudinal control signals with a minimal number of modulation frequencies.

The input mode cleaner must simultaneously stabilize the laser frequency and transmit all control sidebands with minimal attenuation.
To satisfy this requirement, the input mode cleaner round-trip length is chosen as
$
L_{\mathrm{IMC}}
=
10.2~\mathrm{m},
$, $
\mathrm{FSR}_{\mathrm{IMC}}
\simeq
14.8~\mathrm{MHz},$

and an additional modulation frequency is introduced at
$
f_{\mathrm{IMC}}
=
44.2~\mathrm{MHz}.$
The resulting frequency plan ensures compatibility between the input mode cleaner and the main interferometer while preserving the resonance conditions required for cavity control.
All selected sidebands can therefore be transmitted through the  input mode cleaner and utilized for interferometer sensing without compromising laser-frequency stabilization.

The complete sideband configuration adopted in this work follows the design principles established for advanced gravitational-wave detectors and provides a practical control architecture for the CHRONOS prototype interferometer~\cite{Kwee2012,Evans2015}.

%%%%%%%%%%%%%
\subsection{Control ports and control scheme}
\label{subsec:control_ports}

The CHRONOS interferometer employs a dual-recycled Sagnac configuration,
using multiple sensing ports and RF sidebands to control the longitudinal
and angular degrees of freedom of the coupled optical cavities.

The control architecture is based on a set of dedicated reference ports distributed
throughout the interferometer.
The corresponding port locations are indicated by the red labels in
Fig.~\ref{fig:layout}.
These ports provide the error signals required for cavity locking,
alignment control, and balanced-homodyne operation.

This section describes the overall control strategy,
including the initial-alignment phase and the role of each sensing port
in the interferometer locking sequence~\cite{Evans2015,akutsu2021kagra}.

\subsubsection{Initial alignment and control}

The interferometer lock-acquisition sequence begins with the initial-alignment phase.
Its purpose is to establish a stable optical reference before attempting to lock the coupled cavities of the dual-recycled Sagnac interferometer.
During this stage, the positions and orientations of each mirror for each cavity, and input steering mirrors are adjusted to optimize beam pointing and mode matching, thereby stabilizing the spatial mode and phase reference of the injected beam.

The input mode cleaner serves as both a spatial mode filter and a laser-frequency reference~\cite{Kwee2012,akutsu2021kagra}.
It is implemented as a four-mirror bow-tie cavity and is maintained on resonance using the Pound--Drever--Hall technique~\cite{Drever1983,Black2001}.
An RF sideband at
$
f_{\mathrm{IMC}} = 44.2~\mathrm{MHz}
$
is used for phase modulation, and the reflected light is demodulated at the same frequency to generate the cavity-length error signal.

The input mode cleaner is the first cavity to be locked and establishes the frequency and spatial-mode reference for all subsequent control stages.
Once the input mode cleaner is stabilized, the transmitted beam provides a well-defined optical field for the downstream recycling cavities and the Sagnac interferometer.
The input mode cleaner mirrors as indicated with yellow label in Fig.~\ref{fig:layout} , $IMC1$ -- $IMC4$, are suspended and actuated using coil--magnet actuators ~ \cite{akutsu2021kagra}.
This approach provides large control authority at low frequencies while maintaining good mechanical isolation from ground motion.
The transmitted beam is continuously monitored to verify spatial-mode quality and intensity stability, and its low-frequency component is also used for offset compensation in the laser-frequency stabilization loop~\cite{Kwee2012}.

The RF phase reference of the IMC is phase-locked to the modulation sources at
$
f_1 = 29.5~\mathrm{MHz},
$, $
f_2 = 59.0~\mathrm{MHz},
$
ensuring a common demodulation phase for the reflection-port ($REFL_j$) and pick-off-port ($POP_j$) sensing channels, where $j =\{X,Y\}$.

After completion of the initial alignment phase, the interferometer has a stable frequency, phase, and spatial-mode reference.
This provides the starting point for the green pre-lock stage and the subsequent acquisition of the power-recycling cavity and signal-recycling cavity  longitudinal control loops.
The reference ports used throughout the lock-acquisition sequence are indicated by the red labels in Fig.~\ref{fig:layout}.
These labels are referenced throughout the remainder of this section when describing the generation of the corresponding control signals.

%%%%%%%%%%%%%%

\subsubsection{Reflection port}
The $REFL_j$ port detects the light reflected from the input mirror of the power-recycling cavity.  
This port directly monitors the total reflectivity of the interferometer 
and provides the primary error signal for the power-recycling cavity length (PRCL).  
Demodulation of the reflected light at $f_1 = 29.5~\mathrm{MHz}$ 
produces the error signal of power-recycling cavity via the Pound--Drever--Hall technique~\cite{Drever1983,Black2001}.  
In addition, the REFL port serves as an auxiliary monitor 
for the common Sagnac mode (CSAG), 
corresponding to the average optical path length of the CW and CCW beams, 
and contributes to maintaining the overall optical resonance and laser-frequency stability.

\subsubsection{Pick-off port}
The $POP_j$ port is a pick-off of the intracavity field in the power-recycling cavity, 
designed to sense both the common and differential components of the optical field~\cite{Aston2012}, where $j = \{ X,Y \}$.  
Two photodiodes are placed at positions with distinct Gouy phases to spatially separate optical modes:
\begin{itemize}
  \item \textbf{$POP_j-A$ (near field):} Located near the $PRM2_{j}$, 
        sensitive to the CSAG 
        and the PRCL.  
        This signal is mainly used as an auxiliary channel 
        for common path-length and laser-frequency stabilization.
  \item \textbf{$POP_j-B$ (far field):} Located downstream with a Gouy-phase shift of $45^{\circ}$--$90^{\circ}$ 
        relative to $POP_j-A$, 
        providing high sensitivity to the differential Sagnac mode (DSAG).  
\end{itemize}
Both $POP_j-A$and $POP_j-B$ are demodulated at $f_2 = 59.0~\mathrm{MHz}$.  
The Gouy-phase difference enables clear separation of the CSAG, PRCL, 
and DSAG in detection.

\subsubsection{Antisymmetric port }
The $AS_j$ port corresponds to the output (dark) port of the Sagnac interferometer.
When the interferometer is on resonance,
the carrier light ideally remains dark at this port (dark fringe).
The demodulated signal at $f_2 = 59.0~\mathrm{MHz}$
provides the primary error signal for controlling
the signal-recycling cavity length (SRCL)~\cite{Evans2015}.

The $AS_X$ and $AS_Y$ signals
are combined in the OBS
and decomposed into common and differential modes.
The differential mode corresponds to the relative angular motion of the mirrors
and therefore carries the scientific signal of interest.
This differential AS signal is injected into the output mode cleaner,
and the final science signal, denoted as DSAG,
is extracted using balanced-homodyne detection~\cite{fricke2012,vahlbruch2010}.

\subsubsection{Comparison with a Fabry--Perot Michelson interferometer}
The DSAG corresponds to the Differential arm length (DARM) degree of freedom in a Michelson interferometer, 
but in the CHRONOS configuration it represents an \emph{angular momentum}-type observable 
that measures the difference in propagation speed between the CW and CCW beams 
rather than their positional displacement~\cite{chen2003}.  
Similarly, the CSAG corresponds to the Michelson Common Arm Length (CARM), 
representing the common path length coupled to the laser frequency and power-recycling cavity resonance.

\begin{table*}[t]
\centering
\caption{
Major detection ports and their corresponding control degrees of freedom
in the CHRONOS dual-recycled Sagnac interferometer.
The top row, IMC-INI, represents the initial-alignment and frequency-reference stage
in the IMC,
where the IMCL
is stabilized using a 44.2~MHz Pound--Drever--Hall (PDH) signal.
The subsequent ports---REFL, POP, and AS---are used to sense and control
the common and differential DOFs of the main dual-recycled Sagnac interferometer
(PRCL, CSAG, DSAG, and SRCL).
All RF ports employ phase-coherent modulation sidebands
at $f_1 = 29.5~\mathrm{MHz}$ and $f_2 = 59.0~\mathrm{MHz}$,
which correspond to the resonance conditions of the
PRC and the SRC, respectively.
The final port, AS-BHD, is a DC readout channel that extracts
the velocity component of the DSAG
as the speed-meter signal, which constitutes the scientific output of the interferometer.
The physical locations of the individual detection and control ports
are illustrated in Fig.~\ref{fig:layout}.
}
\label{tab:control_ports}

\resizebox{\textwidth}{!}{
\begin{tabular}{lcccl}
\hline\hline
Port & RF freq. & Main DOF & Role & Remark \\
\hline
$IMC-INI$ & 44.2 MHz & IMCL & IMC length and alignment control & Four-mirror suspended bow-tie cavity \\
$REFL_j$   & 29.5 MHz & PRCL, CSAG & PRC and common path-length control & Also used for frequency stabilization \\
$POP_j-A$  & 59.0 MHz & PRCL (aux), CSAG & Common Sagnac mode monitor & Near-field pick-off \\
$POP_j-B$  & 59.0 MHz & DSAG & Differential Sagnac control & Far-field pick-off \\
$AS_j$     & 59.0 MHz & SRCL & Signal-recycling cavity control & Output port (dark fringe) \\
$AS-BHD$ & DC & Science (DSAG velocity) & Speed-meter readout & Phase-quadrature homodyne detection \\
\hline
\end{tabular}
}
\end{table*}

%%%%%%%%%%%%%%%%%%%%%%%%
\section{Mode-matching design}
\label{sec:mode_matching}

Efficient coupling between the input beam, the Sagnac cavity,
and the recycling cavities is essential for achieving the target
sensitivity of CHRONOS.
Mode mismatch reduces the circulating power,
increases prompt reflection,
and degrades the quantum-noise performance of the speed-meter configuration.

The cavity eigenmodes were evaluated using the ABCD-matrix formalism
for Gaussian beams~\cite{Kogelnik1966}.
For a resonant cavity, the fundamental eigenmode is obtained from the
round-trip matrix and described by the complex beam parameter $q$.
Detailed derivations of the Gaussian-beam propagation,
cavity eigenmode calculation, and mode-overlap formalism
are provided in Appendix~\ref{app:mode_matching}.

The mode matching between the injected beam and the cavity eigenmode
is quantified by the overlap integral

\begin{equation}
\eta =
\left|
 \frac{
   \displaystyle \iint E_{\mathrm{in}}^*(x,y)\,
   E_{\mathrm{cav}}(x,y)\,dx\,dy
 }{
   \sqrt{\displaystyle \iint |E_{\mathrm{in}}|^2 dx\,dy}
   \sqrt{\displaystyle \iint |E_{\mathrm{cav}}|^2 dx\,dy}
 }
\right|^2,
\end{equation}

where $E_{\mathrm{in}}$ and $E_{\mathrm{cav}}$ denote the injected beam
and the cavity eigenmode, respectively.
Perfect overlap corresponds to $\eta=1$.

In CHRONOS, the radii of curvature of the recycling mirrors
and the external beam parameters were optimized
to maximize the overlap between the reference Gaussian beam
and the cavity eigenmodes.

The optimized design achieves mode-matching efficiencies exceeding
99.5\% for both the power-recycling cavity and signal-recycling cavity,
satisfying the requirement for low-loss operation of the speed-meter interferometer.

%%%%%%%%%%%%%%%%%%%%
\subsection{Requirement conditions}
\label{subsec:requirement}

Efficient mode matching is required to ensure stable cavity control,
high circulating power, and preservation of the QND
response of the speed-meter interferometer.
For the CHRONOS triangular ring cavity,
the dominant design constraints are the overall mode-matching efficiency
and the coupling symmetry between the CW and CCW
propagation paths.

In the mode-matching design of the CHRONOS triangular ring cavity,
the internal optical loss $L_{\mathrm{int}}$ is sufficiently small---below $10^{-5}$ (10~ppm)---
thanks to high-quality dielectric coatings.

Therefore, the cavity-control performance is primarily determined by the external coupling efficiency $\eta$
rather than by the internal loss.
The impact of optical loss on the detector sensitivity through the achievable cavity finesse
is discussed separately in Sec.~\ref{sec:noise_model}.

The coupling loss is defined as $L_{\mathrm{cpl}} = 1 - \eta.$

If the coupling efficiencies for the CW and CCW beams are denoted by
$\eta_{\mathrm{CW}}$ and $\eta_{\mathrm{CCW}}$, respectively,
their difference $\Delta \eta = \eta_{\mathrm{CW}} - \eta_{\mathrm{CCW}}$

represents the coupling asymmetry.
The relative asymmetry is given by

\begin{equation}
\left|\frac{\Delta \eta}{\eta}\right|=
\left|
\frac{\eta_{\mathrm{CW}}-\eta_{\mathrm{CCW}}}
{(\eta_{\mathrm{CW}}+\eta_{\mathrm{CCW}})/2}
\right|.
\end{equation}

This quantity determines the balance of the CW and CCW circulating powers
and therefore directly affects the cancellation of radiation-pressure noise
in the speed-meter configuration~\cite{chen2003}.

\paragraph{Asymmetry requirement}

The radiation-pressure noise appearing in the output of a speed-meter interferometer
can be approximated as~\cite{chen2003}

\begin{equation}
S_{x,\mathrm{BA}}
\propto
\frac{1}{\mathcal{F}^{2}}
+
\delta^{2},
\end{equation}

where
$\delta
\equiv
\frac{\Delta\eta}{\eta}
$
denotes the CW/CCW coupling asymmetry.

To preserve the QND cancellation of radiation pressure,
$\delta^2
\ll
\frac{1}{\mathcal{F}^2},$
which gives

\begin{equation}
\left|
\frac{\Delta\eta}{\eta}
\right|
\ll
\frac{1}{\mathcal{F}}.
\end{equation}

For the CHRONOS cavity,
$\mathcal{F}
\simeq
1\times10^{4},$
yielding
$
\frac{1}{\mathcal{F}}
\simeq
3.3\times10^{-5},
$
the coupling asymmetry requirement is taken to be
\begin{equation}
\left|
\frac{\Delta \eta}{\eta}
\right|
<
3\times10^{-5}.
\end{equation}

\paragraph{Mode-matching requirement}

In principle, the coupling loss should remain significantly smaller than the internal cavity loss.
However, for modern low-loss optical systems,
alignment tolerances, residual wavefront distortions,
and higher-order spatial modes become the dominant limitations.

Considering these practical constraints,
a mode-matching efficiency of
$
\eta
\ge
99.5\%$

is generally regarded as an achievable and reliable design target
for precision interferometers~\cite{Kwee2012}.

This requirement is sufficient to ensure efficient power buildup,
stable cavity control,
and reproducible operation,
while comfortably satisfying the CW/CCW asymmetry constraint.

Therefore, the design requirements adopted for the CHRONOS optical system are

\begin{equation}
\eta \ge 99.5\%,
\qquad
\left|
\frac{\Delta\eta}{\eta}
\right|
<
3\times10^{-5}. \label{eq:opt_req}
\end{equation}

These criteria were used throughout the optical optimization presented in the following sections.

\subsection{Optimization strategy}

The optical parameters were optimized to satisfy the mode-matching
requirements obtained as Eq.~\eqref{eq:opt_req} 
while maintaining stable cavity operation.
The mirror positions were fixed by the optical layout, an optimization was performed to maximize the
mode-matching efficiency $\eta$ of the cavity,
using the radius of curvature $R$ of each curved mirror as a design variable.

For each set of curvature parameters,
the cavity eigenmodes were calculated using the ABCD-matrix formalism,
and the corresponding beam-radius profiles and mode-matching efficiencies
were evaluated.
The optimization was constrained to preserve the symmetry between
the CW and CCW propagation paths,
thereby minimizing coupling asymmetry and maintaining balanced
radiation-pressure cancellation.
The final curvature parameters were selected by maximizing the
mode-matching efficiency while simultaneously satisfying the cavity
stability condition.
The detailed optimization procedure, including the evaluation of
astigmatism, Gouy phase, transverse-mode spacing,
and Gaussian-beam overlap integrals,
is described in Appendix~\ref{app:optimization}.

%%%%%%%%%%%%%%%%%
\subsection{Optimization results}
The optimization converged to a unique set of mirror curvatures
that simultaneously satisfies the cavity-stability condition
and the mode-matching requirements derived in
Sec.~\ref{subsec:requirement}.
The resulting optical parameters are summarized in
Table~\ref{tab:alloptics}.

The final mode-matching efficiencies at the principal coupling interfaces
are summarized in Table~\ref{tab:optimized_dq_eff}.
At both the left and right input-test-mass planes,
the coupling efficiencies between the power-recycling cavity and ring cavity,
as well as between the signal-recycling cavity and ring cavity,
reach approximately 99.5\%.

As shown in Table~\ref{tab:alloptics} in Appendix~\ref{app:optics},
the optimal radius of curvature $R$ for each mirror was determined.

Under these conditions,
at the input test-mass planes (ITML and ITMR),
the mode-matching efficiency for the PR--Ring and SR--Ring coupling
reached $\eta \sim 99.5\%$,
thereby satisfying the design requirement
Eq.~\eqref{eq:opt_req} .

The CW and CCW eigenmodes were found to be identical within numerical precision,
and the resulting coupling asymmetry remained below the requirement
Eq.~\eqref{eq:opt_req}.
Therefore, the optical configuration preserves the balanced
radiation-pressure cancellation required for speed-meter operation.

The differential beam parameters $\Delta q$,
wavefront-curvature difference $\Delta R$,
beam-radius difference $\Delta w$,
and corresponding efficiencies $\eta$
obtained at each coupling point
are summarized in Table~\ref{tab:optimized_dq_eff}.

The small values of $\Delta q$, $\Delta R$, and $\Delta w$
confirm that the cavity eigenmodes and injected beams
are nearly identical at each coupling interface.
A few-millimeter difference was observed between the sagittal and tangential focal positions,
but this remains within the acceptable range of astigmatism.
The focus positions and corresponding beam-waist radii
are summarized in Table~\ref{tab:focus_beamwidth}.
The optimized configuration exhibits only a small residual astigmatism,
with negligible impact on the coupling efficiency.

These results demonstrate that the power-recycling , signal-recycling, and ring cavities
can be simultaneously mode matched while maintaining the symmetry
required for QND operation.

\subsection{Cavity stability}

To verify that the optimized mode-matching solution corresponds
to a stable optical configuration,
the cavity stability was evaluated using both
ABCD-matrix calculations and \textsc{Finesse3} simulations.

The resulting cavity parameters are summarized in
Table~\ref{tab:ring_stability}.
For all ring cavities, the round-trip Gouy phase satisfies
$
0^\circ < \psi < 180^\circ,
$
corresponding to the stable-cavity condition.
The sagittal and tangential Gouy phases agree to within
approximately $0.6^\circ$,
indicating that only a small residual astigmatism remains after optimization.
Likewise, the transverse-mode spacings of the two planes differ by less than
$0.2~\mathrm{MHz}$,
confirming consistent eigenmode formation in both directions.

The optimized cavity has a finesse of
$\mathcal{F}=3.14\times10^4$
and a free spectral range of
$\mathrm{FSR}=59.0~\mathrm{MHz}$,
corresponding to a linewidth of
$1.88~\mathrm{kHz}$
and a cavity pole frequency of
$939~\mathrm{Hz}$.
These values satisfy the design requirements for optical storage time
and quantum-noise performance.

The left and right ring cavities exhibit nearly identical optical parameters,
demonstrating that the symmetry condition required for balanced CW and CCW propagation
is preserved throughout the interferometer.

Therefore, all ring cavities satisfy the stability condition
and remain well separated from higher-order-mode degeneracies.
The optimized optical configuration provides a stable basis
for the quantum-noise analysis presented in
Sec.~\ref{sec:noise_model}.

%============================================================
\begin{table*}[t]
\centering
\caption{
Optimized mode-matching results between the power-recycling/signal-recycling cavities and the ring cavities
at the $ITM_{ij}$ planes, where $i=\{L, R\}$ and $j=\{ X,Y \}$.  
Listed are the real and imaginary parts of the differential beam parameter
($\Delta q_{\mathrm{re}}$, $\Delta q_{\mathrm{im}}$), its magnitude $|\Delta q|$, 
differences in wavefront curvature ($\Delta R$),
beam radius ($\Delta w$), and the resulting mode-matching efficiency $\eta$ after optimization.
All efficiencies exceed the target threshold of $\eta \ge 99.5\%$.
Positive $\Delta R$ and $\Delta w$ indicate a larger radius of curvature and beam size in the PR/SR modes relative to the ring mode.
}
\label{tab:optimized_dq_eff}
\resizebox{\textwidth}{!}{
\begin{tabular}{lcccccccc}
\hline\hline
Arm & Plane & Pair & $\Delta q_{\mathrm{re}}$ [m] & $\Delta q_{\mathrm{im}}$ [m] & $|\Delta q|$ [m] & $\Delta R$ [m] & $\Delta w$ [m] & $\eta$ [\%] \\
\hline
$ITM_{Lj}$ & s & PR--Ring & $-5.12\times10^{-3}$ & $-6.90\times10^{-2}$ & $6.91\times10^{-2}$ & $-0.82$ & $-2.65\times10^{-5}$ & 99.6 \\
     & s & SR--Ring & $-1.33\times10^{-2}$ & $-7.22\times10^{-2}$ & $7.34\times10^{-2}$ & $-0.20$ & $-2.85\times10^{-5}$ & 99.6 \\
     & t & PR--Ring & $-2.07\times10^{-2}$ & $+6.93\times10^{-2}$ & $7.23\times10^{-2}$ & $+4.31$ & $+2.35\times10^{-5}$ & 99.7 \\
     & t & SR--Ring & $+1.24\times10^{-2}$ & $+9.05\times10^{-2}$ & $9.14\times10^{-2}$ & $+0.64$ & $+3.29\times10^{-5}$ & 99.5 \\
$ITM_{Rj}$& s & PR--Ring & $+7.16\times10^{-3}$ & $-7.62\times10^{-2}$ & $7.66\times10^{-2}$ & $+0.80$ & $-2.94\times10^{-5}$ & 99.6 \\
     & s & SR--Ring & $+1.55\times10^{-2}$ & $-7.92\times10^{-2}$ & $8.07\times10^{-2}$ & $+0.11$ & $-3.13\times10^{-5}$ & 99.5 \\
     & t & PR--Ring & $+2.36\times10^{-2}$ & $+6.21\times10^{-2}$ & $6.64\times10^{-2}$ & $-4.83$ & $+2.07\times10^{-5}$ & 99.7 \\
     & t & SR--Ring & $-9.08\times10^{-3}$ & $+8.39\times10^{-2}$ & $8.44\times10^{-2}$ & $-0.78$ & $+3.03\times10^{-5}$ & 99.6 \\
\hline
\end{tabular}
}
\end{table*}

%============================================================

%============================================================
\begin{table}[h]
\centering
\caption{
Beam waist $w_0$ at each defined focus for sagittal and tangential planes.
The positions of the foci are shown in Fig.~\ref{fig:layout}.
}
\label{tab:focus_beamwidth}
\begin{tabular}{ccc}
\hline\hline
Plane & Focus & $\bar{w}_0$ (mm) \\ \hline
s & $F_{\mathrm{RING_{LY}}}, F_{\mathrm{RING_{RX}}}$ & 0.452 \\
s & $F_{\mathrm{RING_{RY}}}, F_{\mathrm{RING_{LX}}}$ & 0.455 \\
s & $F_{\mathrm{CM_{Y}}}, F_{\mathrm{CM_{X}}}$ & 0.045 \\
s & $F_{\mathrm{SAG_{LY}}}, F_{\mathrm{SAG_{RX}}}$ & 0.452 \\
s & $F_{\mathrm{SAG_{RY}}}, F_{\mathrm{SAG_{LX}}}$ & 0.455 \\
s & $F_{\mathrm{PRM_{Y}}}, F_{\mathrm{PRM_{X}}}$ & 2.617 \\
s & $F_{\mathrm{SRM_{Y}}}, F_{\mathrm{SRM_{X}}}$ & 2.640 \\ \hline
t & $F_{\mathrm{RING_{LY}}}, F_{\mathrm{RING_{RX}}}$ & 0.455 \\
t & $F_{\mathrm{RING_{RY}}}, F_{\mathrm{RING_{LX}}}$ & 0.457 \\
t & $F_{\mathrm{CM_{Y}}}, F_{\mathrm{CM_{X}}}$ & 0.046 \\
t & $F_{\mathrm{SAG_{LY}}}, F_{\mathrm{SAG_{RX}}}$ & 0.455 \\
t & $F_{\mathrm{SAG_{RY}}}, F_{\mathrm{SAG_{LX}}}$ & 0.457 \\
t & $F_{\mathrm{PRM_{Y}}}, F_{\mathrm{PRM_{X}}}$ & 2.604 \\
t & $F_{\mathrm{SRM_{Y}}}, F_{\mathrm{SRM_{X}}}$ & 2.630 \\
\hline\hline
\end{tabular}
\end{table}
%============================================================

%============================================================
\begin{table}[h]
\centering
\caption{
Simulated stability and optical parameters of the left and right ring cavities after optimization. TMS and FSR are the transverse-mode spacing and free spectral range.
}
\label{tab:ring_stability}
\begin{tabular}{lccccc}
\hline\hline
Cavity & Plane & $\psi$ [deg] &TSR [MHz] & FSR [MHz] & Pole [Hz] \\ \hline
Left ring  & s & 153.3 & 50.2 & 59.0 & 939 \\
Left ring  & t & 153.0 & 50.1 & 59.0 & 939 \\
Right ring & s & 153.0 & 50.1 & 59.0 & 939 \\
Right ring & t & 152.7 & 50.0 & 59.0 & 939 \\
\hline\hline
\end{tabular}
\end{table}

%============================================================
\section{Sensitivity} \label{sec:sensitivity}

The sensitivity of the CHRONOS interferometer
is determined by a combination of quantum noise,
thermal noise, seismic disturbances,
Newtonian noise, and technical noise sources.
Among these contributions,
quantum noise plays a central role in defining the performance of
the Sagnac speed-meter topology,
which is the primary focus of this work.

The optical and mechanical parameters used in the sensitivity analysis
are summarized in Table~\ref{tab:params}.
These parameters define the reference operating point adopted throughout
this paper and were selected to provide near-optimal sensitivity
in the target observation band around $0.1$--$1~\mathrm{Hz}$.

In the first part of this section,
we introduce the quantum-noise model,
including the input--output relations,
effective optomechanical coupling,
and the influence of the homodyne detection angle and cavity detuning.
In the latter part, numerical simulations using \textsc{Finesse3}~\cite{Brown2025Finesse}
are presented to show the resulting sensitivity curves,
focusing on the effects of End test mass reflectance, power-recycling cavity detuning, signal-recycling cavity detuning,
and homodyne detection angle.

The detailed models of thermal, seismic,
Newtonian, and other technical noise sources
are summarized in Appendix~\ref{app:noise},
while the intensity-noise analysis is presented separately in
Tanabe \textit{et al.}~\cite{Tanabe_2026}.
The Newtonian noise analysis of CHRONOS is summarized in {\it Onglao et al.}~\cite{onglao2026prospectsobservinggravitygradientnoise}.

This approach allows the role of the speed-meter quantum-noise suppression
to be isolated and evaluated while retaining a realistic estimate of the
full detector sensitivity.

\begin{table}[t]
  \centering
  \caption{Assumed optimal optical and mechanical parameters for the CHRONOS interferometer. In the CHRONOS phase convention, the detuning phase of the power-recycling mirror, signal-recycling and the homodyne detection phase are referenced to the $90^{\circ}$ baseline. This convention is consistently used throughout the detuning and quadrature-rotation analysis.}
  \label{tab:params}
  \begin{tabular}{l|c|c}
    \hline\hline
    Definition & Symbol & Value \\ \hline
    Signal-recycling mirror reflectivity & $R_s$ & 0.5 \\
    Power-recycling mirror reflectivity & $R_p$ & 0.9 \\
    Input test-mass reflectivity & $R_i$ & 0.9999 \\
    Input laser power & $P_{\mathrm{in}}$ & $1~\mathrm{W}$ \\
    Circulating power in arm cavity & $P_{\mathrm{arm}}$ & $444~\mathrm{W}$ \\
    PRC detuning phase & $\phi_p$ & $5^{\circ}$ \\
    SRC detuning phase & $\phi_s$ & $0^{\circ}$ \\
    Homodyne detection angle & $\zeta$ & $46^{\circ}$ \\
    Ring-cavity finesse & $\mathcal{F}$ & $3.14\times10^4$ \\
    Beam radius on end test mass & $w$ & $2.6~\mathrm{mm}$ \\ \hline
   Mass of end test mass & $M_{\mathrm{ETM}}$ & $171~\mathrm{kg}$ \\
    Moment of inertia of test mass & $I$ & $19.9~\mathrm{kg\,m^2}$ \\
    Torsion-bar length & $L_{\mathrm{bar}}$ & $1~\mathrm{m}$ \\
    Geometrical coupling factor & $\eta$ & 0.936 \\ \hline\hline
  \end{tabular}
\end{table}

\subsection{Noise model}
\label{sec:noise_model}

The dominant fundamental noise source in CHRONOS is quantum noise,
which arises from vacuum fluctuations entering the interferometer.
As in conventional laser interferometers,
quantum noise consists of shot noise and radiation-pressure noise.
However, unlike a Michelson position meter,
the Sagnac topology measures the velocity of the test masses,
leading to a suppression of quantum back-action at low frequencies.
To quantify this effect,
we employ the input--output formalism developed for Sagnac speed meters.

The interferometer's quantum noise is expressed in terms of the measured angular displacement $\theta$.
Following the Buonanno--Chen formalism~\cite{Buonanno2003}, we use the two-photon representation~\cite{Caves1981}.
In this method, we define the input and output quadrature vectors as
\begin{align}
\boldsymbol{a} &\equiv 
\begin{pmatrix}
a_1 \\[2pt] a_2
\end{pmatrix},
\qquad
\boldsymbol{b} \equiv 
\begin{pmatrix}
b_1 \\[2pt] b_2
\end{pmatrix}.
\end{align}

Here, the input quadrature vector of the interferometer is denoted by $\boldsymbol{a}$.
The amplitude quadrature (in-phase component) $a_1$
and the phase quadrature (out-of-phase component) $a_2$ are represented.
The corresponding output quadrature vector $\boldsymbol{b}$ is also defined.
With these definitions, the input--output relation can be written as
\begin{equation}
\boldsymbol{b}
= \frac{1}{M}\biggl[
  e^{2i(\beta_{\mathrm{sag}}+\Phi_s)} C \boldsymbol{a}
  + \sqrt{2\mathcal{K}_{\mathrm{sag}}}\, t_s e^{i(\beta_{\mathrm{sag}}+\Phi_s)}
  \boldsymbol{D}\frac{\theta}{\theta_{\mathrm{SQL}}}
\biggr],
\label{eq:IOrelation}
\end{equation}
where the transmittance of the signal-recycling mirror is $t_s = \sqrt{1-R_s}$ and its phase delay is $\Phi_s$.
We define $G = C C^{\mathsf{T}}$, and  
$Q = \Re(\boldsymbol{D})\,\Re(\boldsymbol{D})^{\mathsf{T}} + \Im(\boldsymbol{D})\,\Im(\boldsymbol{D})^{\mathsf{T}}$,
where the matrix $C = [c_{ij}]$ and the column vector $\boldsymbol{D}^{\mathsf{T}}=(D_1, D_2)$ are defined following Eqs.~(2.22)--(2.24) of Ref.~\cite{BuonannoChen2001}. 

For a torsion-bar interferometer,
the quantum-noise performance is conveniently normalized
by the standard quantum limit,
which represents the minimum measurement noise achievable
when measurement imprecision and quantum back-action contribute equally.
The standard quantum limit~\cite{Ando2010} for the angular displacement sensitivity of a torsion-bar detector is
\begin{equation}
    \theta_{\rm SQL}(\Omega)
    =
    \sqrt{\frac{2\hbar}{I\Omega^2}},
\end{equation}
where $\theta_{\rm SQL}(\Omega)$ represents the amplitude spectral density of the angular displacement in units of ${\rm rad}/\sqrt{\rm Hz}$.
The phase accumulated in the arm cavity is
\[
    \beta(\Omega) = \arctan\!\left(\frac{\Omega}{\gamma}\right), \qquad
    \beta_{\mathrm{sag}}(\Omega) = 2\beta(\Omega) + \frac{\pi}{2}.
\]
The suppression of quantum back-action is described by the effective
optomechanical coupling factor $\mathcal{K}_{\rm sag}(\Omega)$.
For a Sagnac interferometer,
the coupling is modified by both the velocity-sensitive response
and the resonant enhancement provided by the power-recycling cavity.

The Sagnac-topology optomechanical coupling becomes
\begin{equation}
    \mathcal{K}_{\mathrm{sag}}(\Omega) =
    4\, \mathcal{K}(\Omega)\, |H_{\mathrm{PRC}}(\Omega)|^2
    \sin^2 \beta_{\mathrm{sag}}(\Omega),
\end{equation}
where the frequency-dependent coupling $\mathcal{K}(\Omega)$ and cavity bandwidth $\gamma$
follow the definitions in Ref.~\cite{BuonannoChen2001}.
The transfer function of power-recycling cavity is
\[
    H_{\mathrm{PRC}}(\Omega) =
    \frac{\sqrt{1-R_p}}{1 - \sqrt{R_p}\, e^{2 i \tau_p \Omega} r_{\mathrm{ifo}}},
\]
and the effective interferometer reflectivity (including Sagnac propagation) is
\[
    r_{\mathrm{ifo}} = i |r_{\mathrm{loop}}| \sin(\beta_{\mathrm{sag}} + \tau_{\mathrm{sag}}\Omega),
\]
where the one-way propagation time in the Sagnac loop $\tau_{\mathrm{sag}}$ and the effective reflectance in the ring cavities and Sagnac loop $r_{\mathrm{loop}} \sim r_i$ are defined.

Projecting the output field onto the homodyne quadrature
specified by the detection angle $\zeta$,
the quantum-noise spectral density can be expressed as

\begin{equation}
    \theta(\Omega) =
    \frac{\theta_{\mathrm{SQL}}(\Omega)}{\sqrt{2 \mathcal{K}_{\mathrm{sag}}(\Omega)(1-r_s^2)}}
    \sqrt{
        \frac{\boldsymbol{\nu}^{\mathsf{T}} G \boldsymbol{\nu}}
             {\boldsymbol{\nu}^{\mathsf{T}} Q \boldsymbol{\nu}}
    },
\label{eq:theta_noise}
\end{equation}
where the homodyne detection vector $\boldsymbol{\nu}^{\mathsf{T}}=(\cos\zeta,\,\sin\zeta)$ is defined.
Diagonalization yields an effective coupling constant $\mathcal{K}_{\mathrm{eff}}$ and an effective homodyne angle $\zeta_{\mathrm{eff}}$.
The effective homodyne angle is obtained as
$\zeta_{\mathrm{eff}} = \zeta - \psi(f) + \theta(f)$,
where $\psi(f)$ and $\theta(f)$ denote the principal-axis rotations of $G$ and $Q$, respectively.

While the analytical model provides physical insight into the role of
the optomechanical coupling and homodyne readout,
the sensitivity of the complete interferometer was evaluated using
\textsc{Finesse3}~\cite{Brown2025Finesse}.
The analytical model includes the full optical configuration,
including the Sagnac interferometer, power-recycling cavity,
signal-recycling cavity, and balanced homodyne readout as listed in Table.~\ref{tab:params}.

In the following sections,
the quantum-noise spectrum obtained with \textsc{Finesse3}~\cite{Brown2025Finesse}
is decomposed into shot-noise and radiation-pressure-noise contributions.
This decomposition allows the impact of the speed-meter response,
cavity detuning, and homodyne angle optimization
to be evaluated separately before constructing the total sensitivity budget.

%%%%%%%%%%%%%%%

\subsection{Optical Absorption and Mechanical Loss in Coatings}
\label{subsec:coat_loss}

The optical coatings of the end test masses play a decisive role in determining both the quantum-noise and thermal-noise performance of the interferometer.
To achieve the target sensitivity of CHRONOS in the sub-hertz band,
two coating properties must be simultaneously optimized:
optical absorption and mechanical loss.
Optical absorption reduces the effective cavity finesse and circulating optical power,
while mechanical loss directly contributes to coating Brownian thermal noise.

Recent advances in crystalline mirror coatings have significantly improved the prospects for cryogenic precision interferometry by reducing coating thermal noise, which has become one of the fundamental limitations of ultrastable optical cavities. Previous studies demonstrated that substrate-transferred AlGaAs crystalline coatings simultaneously achieve ultra-low mechanical dissipation and sufficiently low optical loss in cryogenic silicon cavities, overcoming the performance limitations of conventional amorphous dielectric coatings~\cite{Lee2025PRL}. In particular, a fractional frequency instability of $2.5\times10^{-17}$ at an averaging time of 10~s was achieved, corresponding to more than a tenfold reduction in the coating mechanical loss factor. These results establish crystalline coatings as a promising technology for next-generation cryogenic optical systems operating at the $10^{-18}$ stability level. Such developments are directly relevant to gravitational-wave detectors such as CHRONOS, where minimizing coating thermal noise is essential for realizing the target sensitivity in the sub-Hz frequency band.

In CHRONOS, suppression of optical absorption is pursued through the development of advanced low-loss mirror coatings. Plasma-enhanced chemical vapor deposition (PECVD) SiN and SiON multilayers have demonstrated mechanical-loss angles approaching $10^{-5}$ at 100~Hz and room temperature~\cite{Pan2018_SiN_Chao,Chao2023_SiON}. However, their optical absorption remains higher than desired for sub-hertz operation. To address this issue, low-pressure chemical vapor deposition (LPCVD) coatings are being investigated as a promising alternative, offering the potential to reduce optical absorption while preserving low mechanical loss. A detailed study of LPCVD absorption properties is presented in V.~Kumar \textit{et al.}~\cite{Vivek:inprep}. For the present design study, low-loss SiN/SiON coating technologies are adopted as a realistic baseline, while ongoing LPCVD developments are expected to further improve optical performance. Future implementation of crystalline coatings could provide an additional reduction in coating thermal noise while maintaining the high reflectivity required for quantum-noise-limited operation.

Such developments are particularly relevant for future cryogenic gravitational-wave detectors,
where coating thermal noise is expected to become an important sensitivity limitation.
Although further work is required to scale crystalline coatings to larger optics,
the demonstrated optical-loss and scattering performance already satisfies requirements comparable to those considered in this study.
Furthermore, because the beam radius required for CHRONOS is substantially smaller than that of kilometer-scale interferometers such as LIGO,
the current limitations associated with large-aperture crystalline mirrors are less severe.
Crystalline mirror technologies therefore represent a promising candidate for future upgrades of CHRONOS and other cryogenic sub-Hz gravitational-wave observatories.

\paragraph{Optical absorption}

Even ppm-level absorption leads to increased intracavity loss,
reducing the effective reflectivity of the recycling cavities
and lowering the circulating optical power~\cite{Hello1990}.
Because both shot noise and radiation-pressure noise scale with optical power,
quantum noise is highly sensitive to absorption.
An increase in absorption simultaneously raises the shot-noise floor
and weakens the radiation-pressure-noise suppression mechanism,
thereby modifying the overall quantum-noise spectrum.

\begin{figure}[t]
  \centering
  \includegraphics[width=0.8\textwidth]{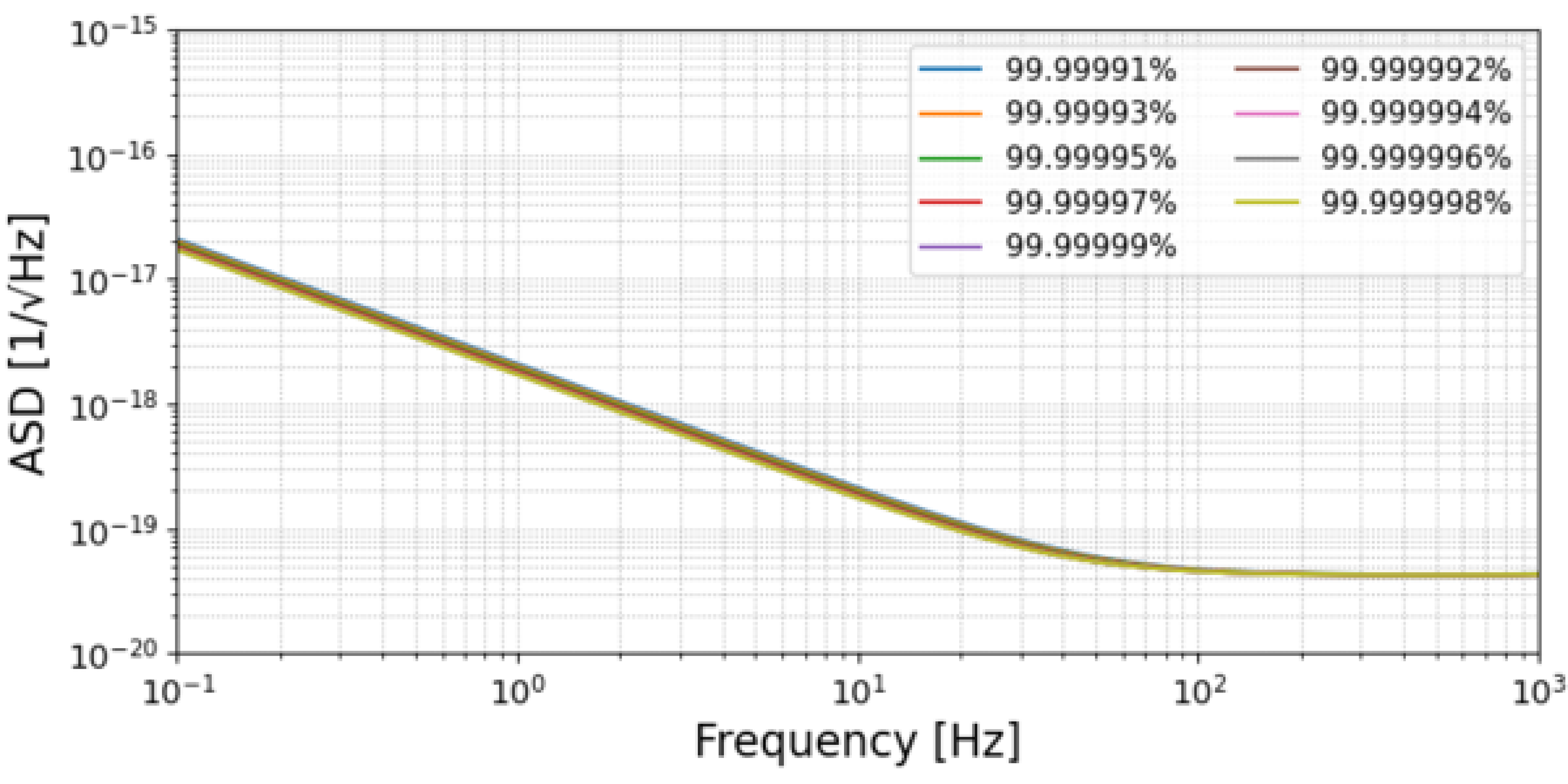}
  \caption{
  Impact of shot noise for varying end test mass reflectivity.
  }
  \label{fig:C_ASD_S}
\end{figure}

\begin{figure}[t]
  \centering
  \includegraphics[width=0.8\textwidth]{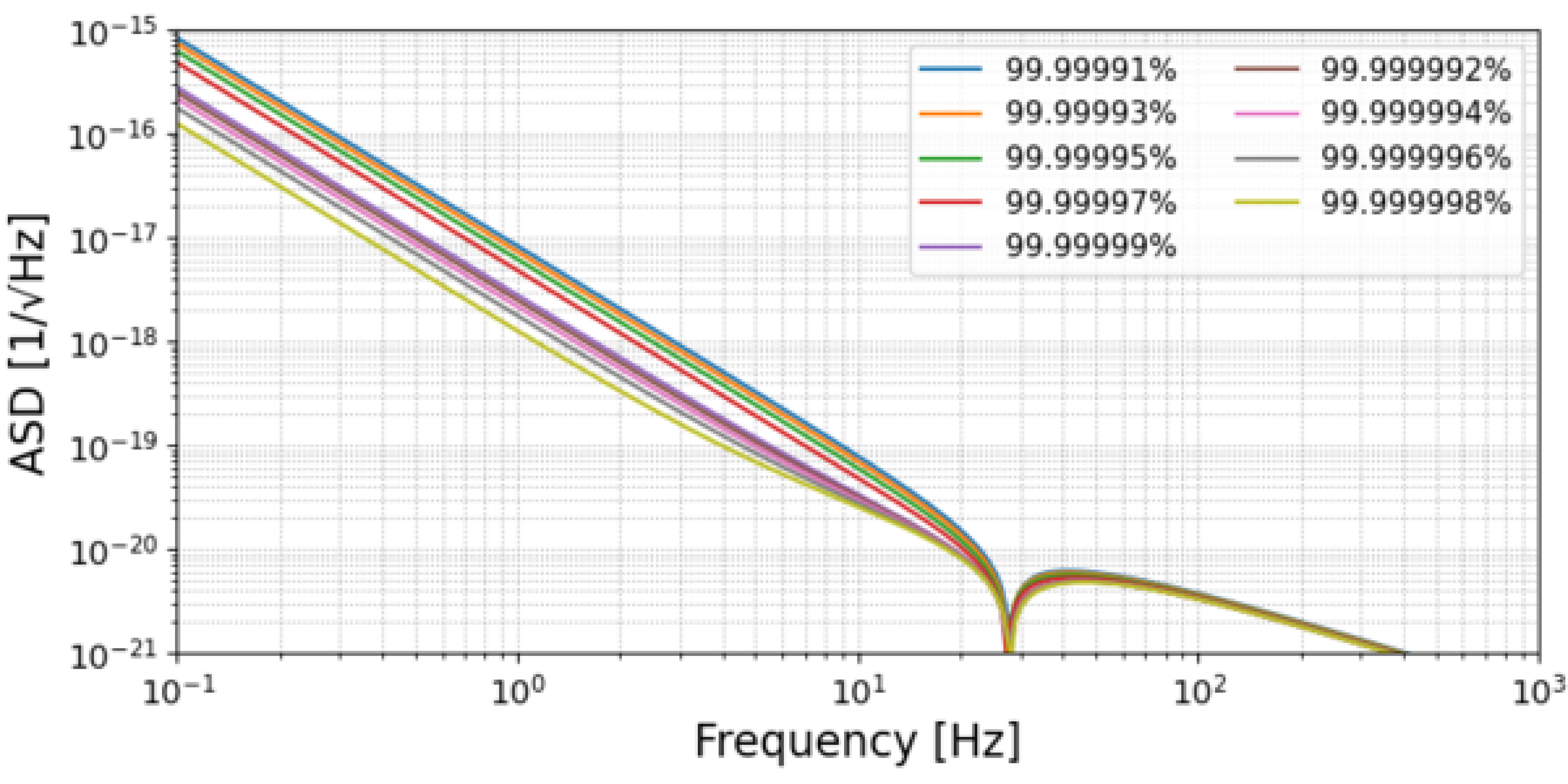}
  \caption{
  Radiation-pressure noise at low frequencies, illustrating degradation of QND suppression with reduced $r_{\mathrm{ifo}}$.
  }
  \label{fig:C_ASD_R}
\end{figure}

Figures~\ref{fig:C_ASD_S} and~\ref{fig:C_ASD_R}
compare the shot-noise and radiation-pressure-noise spectra for different end test mass coating reflectivities.
The variation in shot noise is relatively small because it depends primarily on the circulating optical power.
In contrast, radiation-pressure noise increases significantly at low frequencies as the effective interferometer reflectivity $r_{\mathrm{ifo}}$ is reduced.
As shown in Fig.~\ref{fig:C_ASD_R},
lower reflectivity degrades the quantum back-action cancellation provided by the Sagnac speed-meter topology,
leading to a deterioration of the low-frequency sensitivity.

\paragraph{Mechanical loss}

The mechanical-loss angle of the coating layers directly determines
the amplitude of Brownian thermal fluctuations~\cite{Levin1998,Harry2002}.
Even under cryogenic operation,
coating thermal noise remains an important design consideration for achieving the target CHRONOS sensitivity.

\begin{figure}[t]
  \centering
  \includegraphics[width=0.8\textwidth]{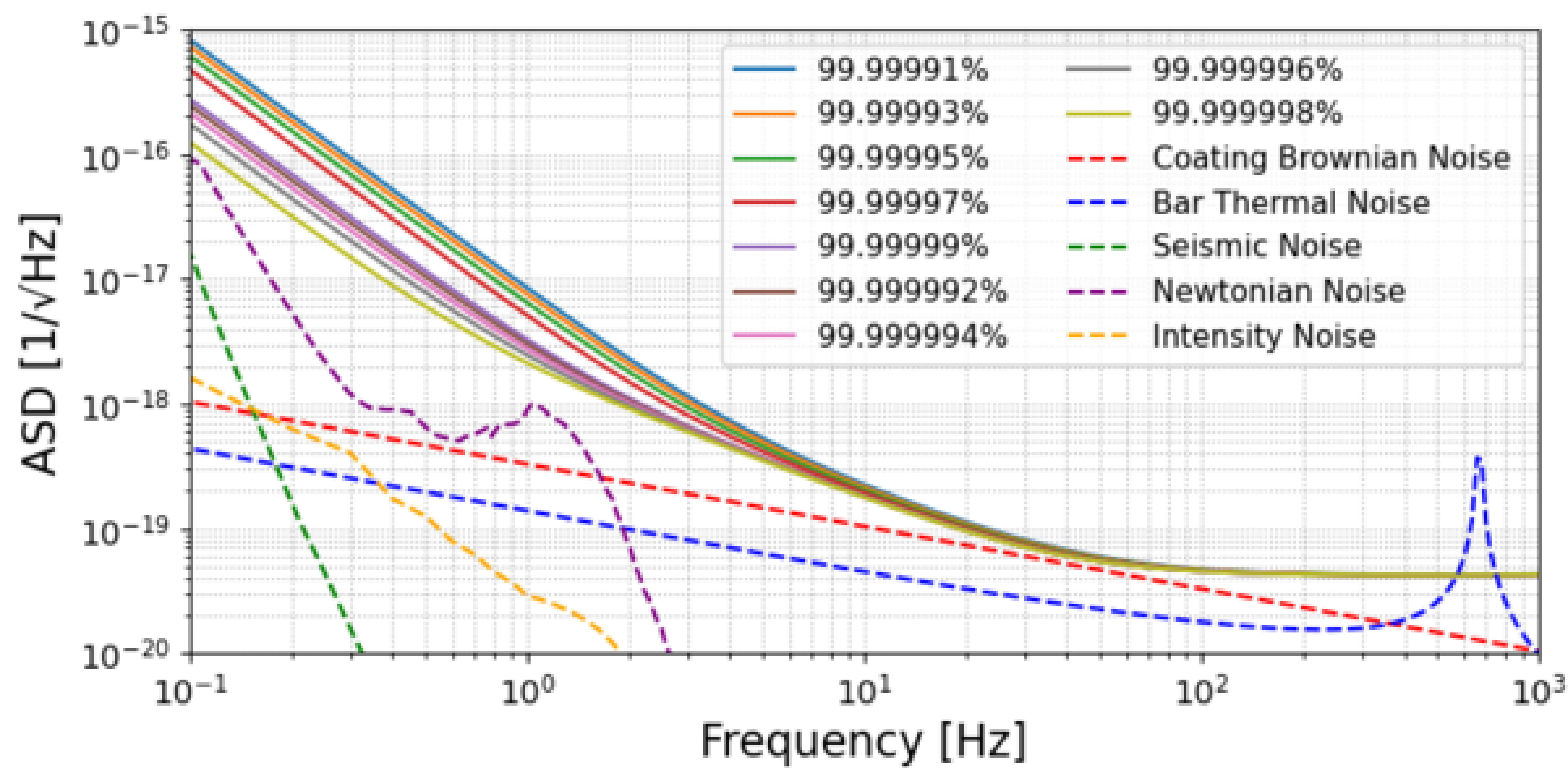}
  \caption{
  Total quantum-noise spectra including shot noise, radiation-pressure noise, and other technical noise sources. 
  Coating effects dominate the sensitivity in the 0.1--10~Hz band. 
  The intensity noise is calculated elsewhere~\cite{Tanabe_2026}.
  }
  \label{fig:C_ASD_TOT}
\end{figure}

Figure~\ref{fig:C_ASD_TOT}
shows the resulting total sensitivity, including shot noise,
radiation-pressure noise, and other technical noise sources.
Although seismic and environmental disturbances dominate at the lowest frequencies,
coating performance remains important because it determines both the achievable cavity finesse and the thermal-noise floor.
The figure demonstrates that maintaining extremely low optical absorption and mechanical loss is essential for preserving the full benefit of the Sagnac speed-meter topology.

To achieve the CHRONOS sensitivity goal,
the total effective coating loss angle should satisfy
$\phi_c^{\mathrm{eff}} < 10^{-5}$
at cryogenic temperatures,
while maintaining a reflectivity of
$R \sim 99.9999\%$.
This dual optimization---simultaneous reduction of optical absorption and mechanical loss---is essential for sustaining the high finesse
($\mathcal{F} \simeq 3.1\times10^{4}$)
and quantum-noise-limited performance required for the CHRONOS sub-hertz interferometer.

%%%%%%%%%%%%%%%%%%
\subsection{Impact of Power-Recycling Cavity Detuning}

The power-recycling cavity plays a central role in determining the quantum-noise performance of CHRONOS.
In addition to increasing the circulating optical power,
the power-recycling cavity modifies the effective optomechanical coupling and therefore influences the balance between shot noise and radiation-pressure noise.
Unlike conventional interferometers, where sensitivity shaping is primarily achieved through signal-recycling detuning,
CHRONOS utilizes power-recycling cavity detuning as a key parameter for optimizing low-frequency performance.
In this work, the reference phase of the power-recycling cavity is defined such that resonance corresponds to a detuning angle of $90^\circ$.
The power-recycling cavity detuning angle quoted throughout this paper is therefore defined as the deviation from this $90^\circ$ reference point.

When the power-recycling cavity is detuned by a phase $\phi_p$,
the transfer function $H_{\mathrm{PRC}}$ is modified,
leading to a change in the effective optomechanical coupling coefficient $\mathcal{K}_{\mathrm{sag}}$.
Physically, this detuning alters the balance between the amplitude and phase quadratures of the circulating field,
thereby changing the quantum-noise correlations responsible for back-action suppression.
As a result, both the shot-noise and radiation-pressure-noise spectra are modified.

\begin{figure}[t]
  \centering
  \includegraphics[width=0.8\textwidth]{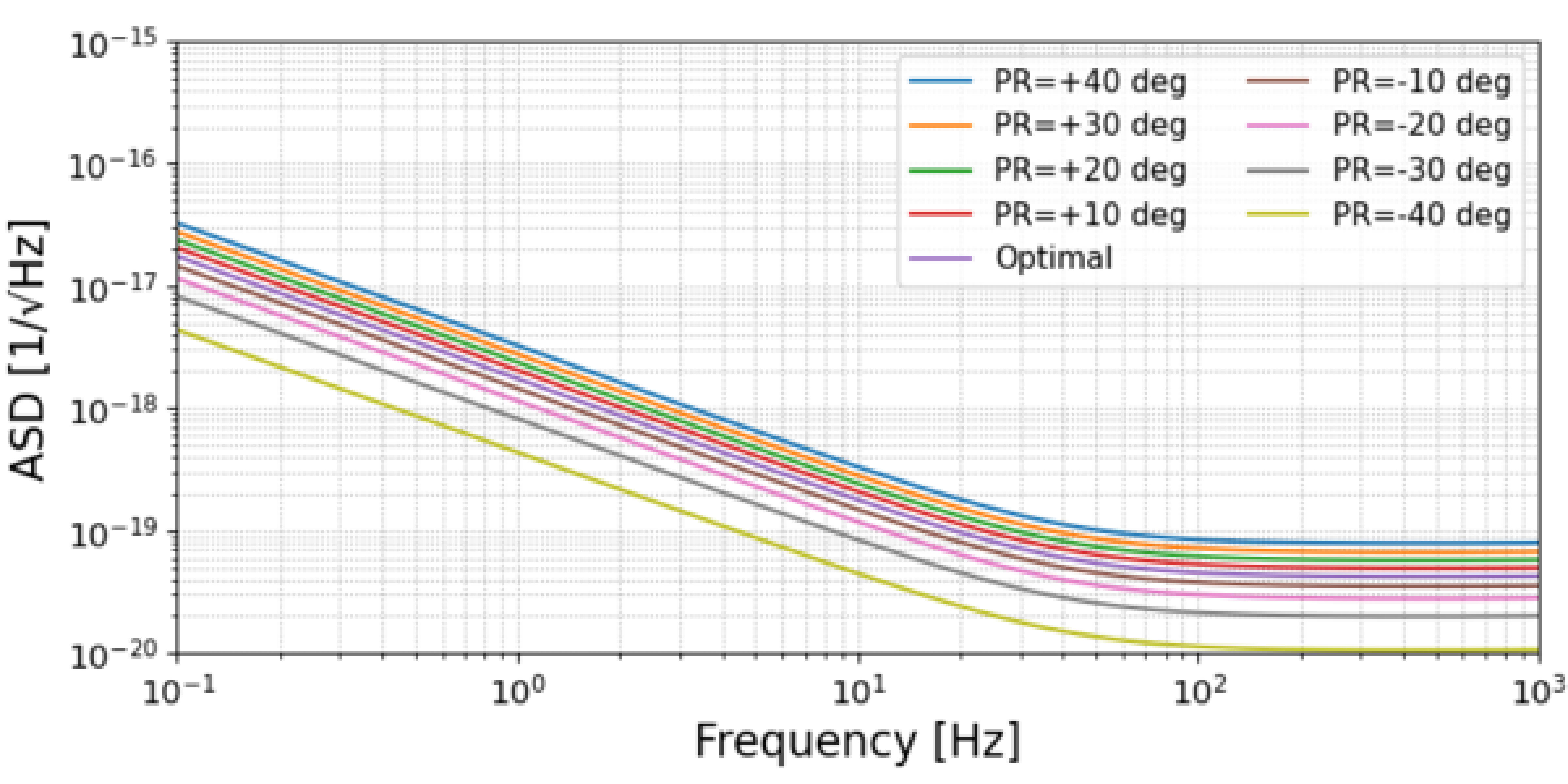}
  \caption{
  Impact of power-recycling cavity detuning on shot noise. 
  The variation is explained by the change in circulating arm power relative to the optimal point.
  }
  \label{fig:PR_ASD_S}
\end{figure}

Figure~\ref{fig:PR_ASD_S} shows the effect of power-recycling cavity detuning on the shot-noise spectrum.
Because shot noise is primarily determined by the circulating optical power,
detuning shifts the cavity away from its optimum operating point and produces a nearly uniform change in the high-frequency noise floor.
The dependence is relatively weak, indicating that shot noise is only moderately affected by power-recycling cavity detuning.

\begin{figure}[t]
  \centering
  \includegraphics[width=0.8\textwidth]{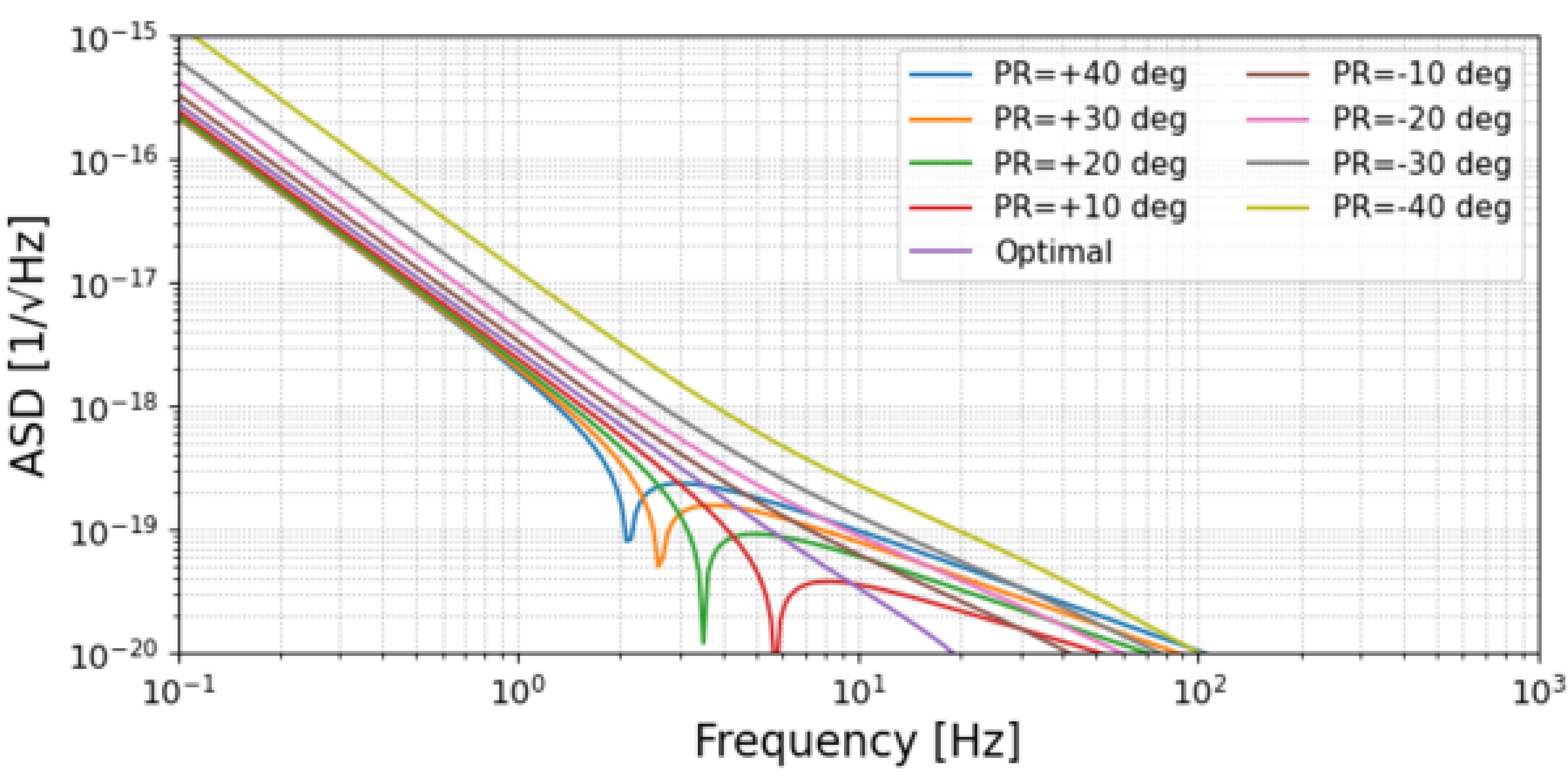}
  \caption{
  Impact of power-recycling cavity detuning on radiation-pressure noise. 
  At the optimal point, back-action correlations and carrier-induced radiation pressure balance with shot noise to yield the best sensitivity near 1~Hz.
  For the present design, the optimal power-recycling cavity detuning angle is $5^\circ$.
  }
  \label{fig:PR_ASD_R}
\end{figure}

The corresponding radiation-pressure-noise spectra are shown in Fig.~\ref{fig:PR_ASD_R}.
In contrast to shot noise,
the low-frequency radiation-pressure-noise contribution exhibits a much stronger dependence on detuning.
Both positive and negative detuning weaken the quantum back-action suppression provided by the Sagnac speed-meter topology,
leading to a substantial increase in low-frequency noise.
This behavior demonstrates that  power-recycling cavity detuning directly controls the strength of the speed-meter response through the effective coupling coefficient $\mathcal{K}_{\mathrm{sag}}$.

\begin{figure}[t]
  \centering
  \includegraphics[width=0.8\textwidth]{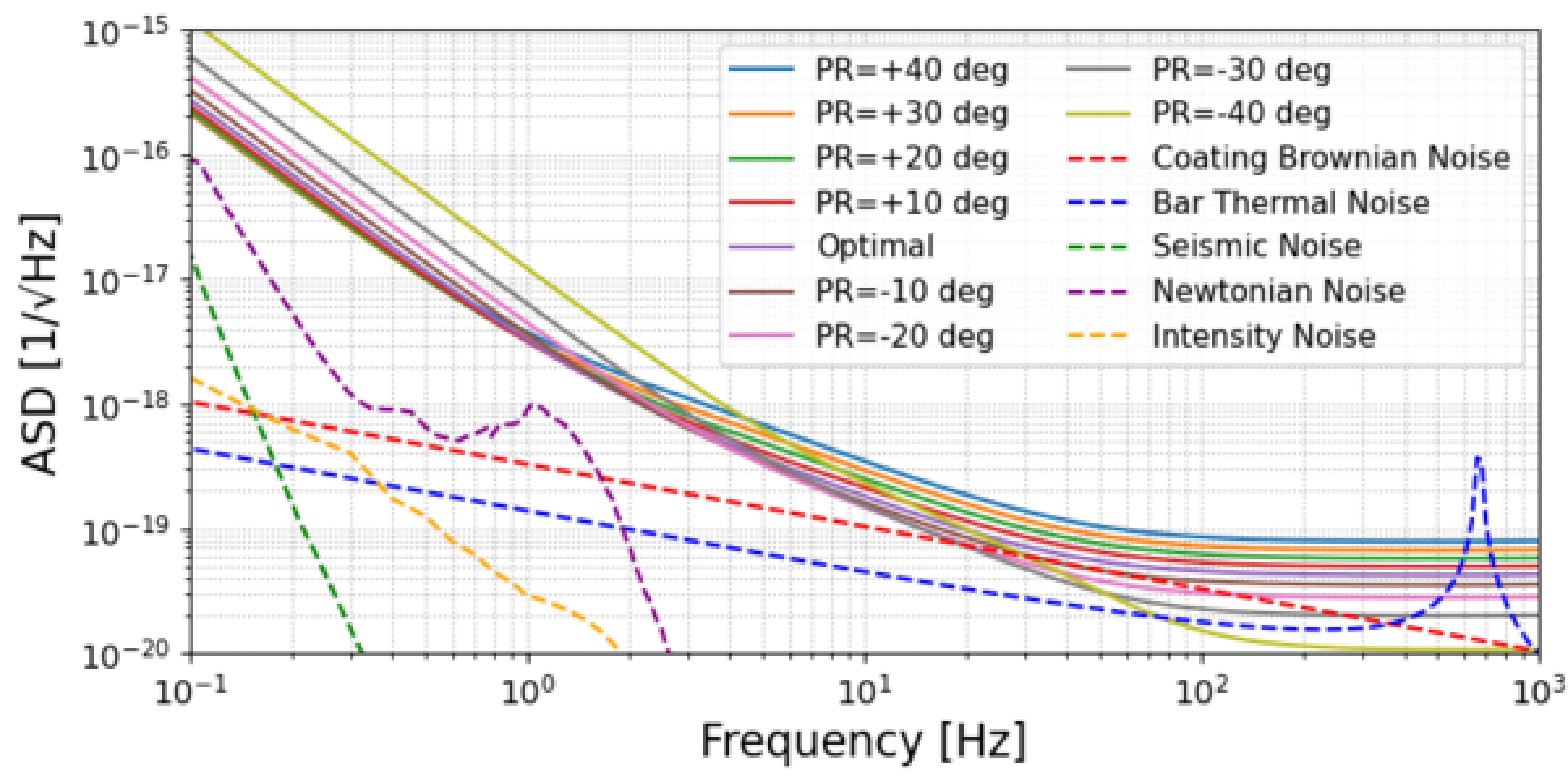}
  \caption{
  Total quantum-noise spectra including shot noise, radiation-pressure noise, and technical noise sources. 
  The optimal point balances the three contributions, while detuning leads to degraded sensitivity. 
  The intensity noise is calculated elsewhere~\cite{Tanabe_2026}. For the present design, the optimal power-recycling cavity detuning angle is $5^\circ$.
  }
  \label{fig:PR_ASD_TOT}
\end{figure}

Figure~\ref{fig:PR_ASD_TOT} presents the resulting total sensitivity,
including shot noise, radiation-pressure noise, and other technical noise sources.
{\color{black}
At the optimum operating point,
the balance between shot noise and radiation-pressure noise maximizes the sensitivity in the target observation band around $0.1$--$1~\mathrm{Hz}$.
}
Away from this optimum,
either radiation-pressure noise at low frequencies or shot noise at high frequencies becomes dominant,
resulting in degraded sensitivity.

These results highlight a distinctive feature of CHRONOS.
Whereas conventional Michelson interferometers typically employ signal-recycling detuning to shape their quantum-noise response~\cite{BuonannoChen2001},
the CHRONOS speed-meter configuration achieves its primary sensitivity optimization through power-recycling detuning.
 power-recycling cavity detuning therefore serves not only as a power-enhancement mechanism,
but also as an effective tool for controlling quantum-noise correlations and optimizing detector performance in the sub-hertz band.

%%%%%%%%%%%%%%%%%%%%%%
\subsection{Impact of Signal-Recycling Cavity Detuning}
\label{sec:impact_SRCD}

The signal-recycling mirror modifies the interferometer response by recycling the gravitational-wave signal sidebands and is widely used in Michelson interferometers to shape the quantum-noise spectrum~\cite{BuonannoChen2001}.
In conventional signal-recycled detectors, detuning the signal-recycling cavity can redistribute sensitivity between low and high frequencies.
However, the behavior of a Sagnac speed-meter interferometer is fundamentally different.
For CHRONOS, detuning the signal-recycling cavity away from resonance generally degrades the quantum-noise performance because it weakens the quantum correlations responsible for back-action suppression.

\begin{figure}[t]
  \centering
  \includegraphics[width=0.8\textwidth]{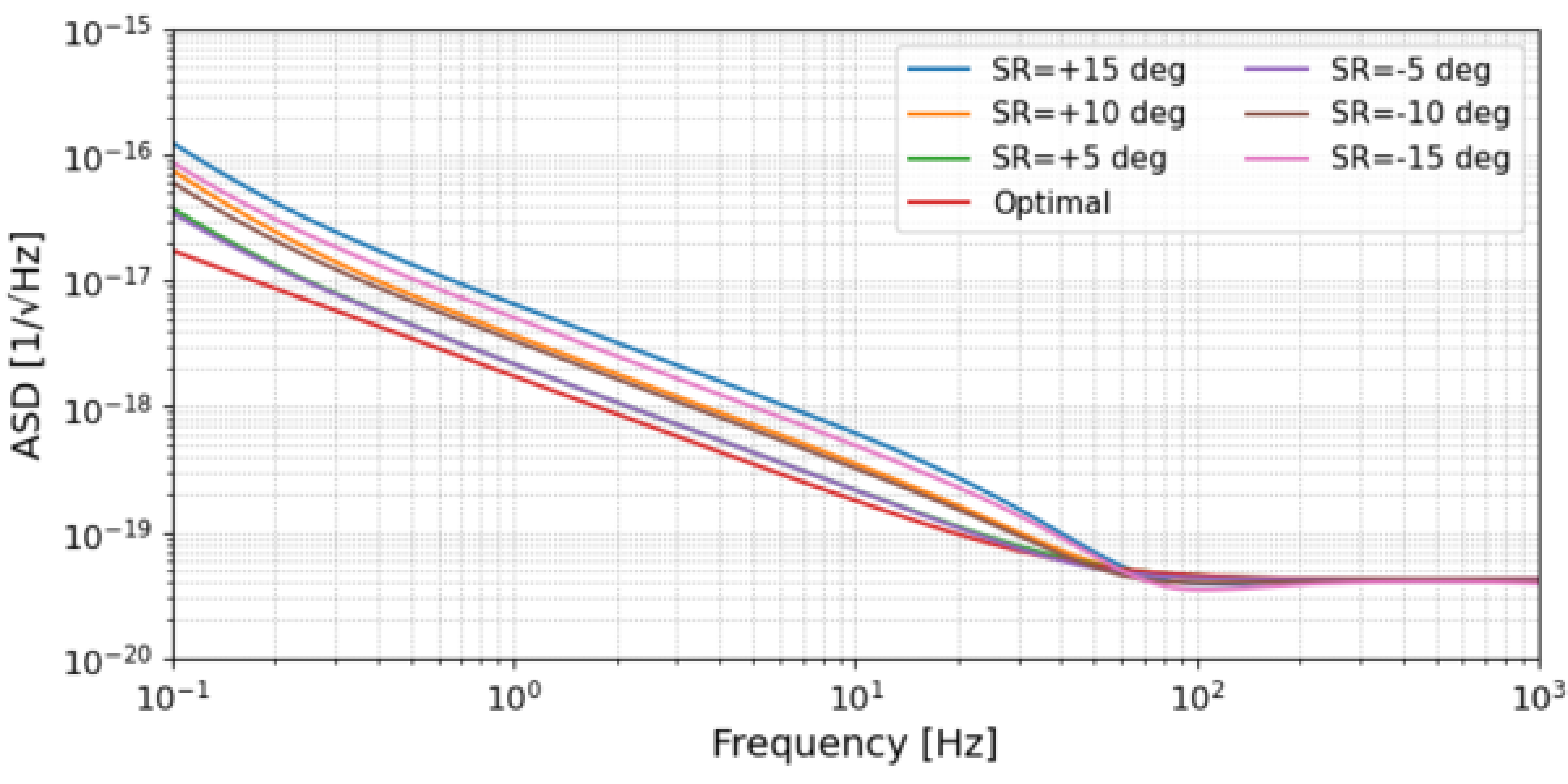}
  \caption{
  Impact of signal-recycling mirror detuning on shot noise.  
  Detuning mixes amplitude and phase quadratures, introducing low-frequency contributions in addition to high-frequency degradation.
  The optimal detuning angle is 0 degree in this plot.
  }
  \label{fig:SR_ASD_S}
\end{figure}

Figure~\ref{fig:SR_ASD_S} shows the effect of signal-recycling cavity detuning on the shot-noise spectrum.
At resonance, the signal sidebands are optimally coupled to the output port, yielding the lowest shot-noise level.
Detuning the signal-recycling cavity mixes the amplitude and phase quadratures in the two-photon formalism,
effectively rotating the measured quadrature basis.
As a result, the shot-noise spectrum shifts almost uniformly across the observation band without providing a significant sensitivity improvement in any particular frequency region.

\begin{figure}[t]
  \centering
  \includegraphics[width=0.8\textwidth]{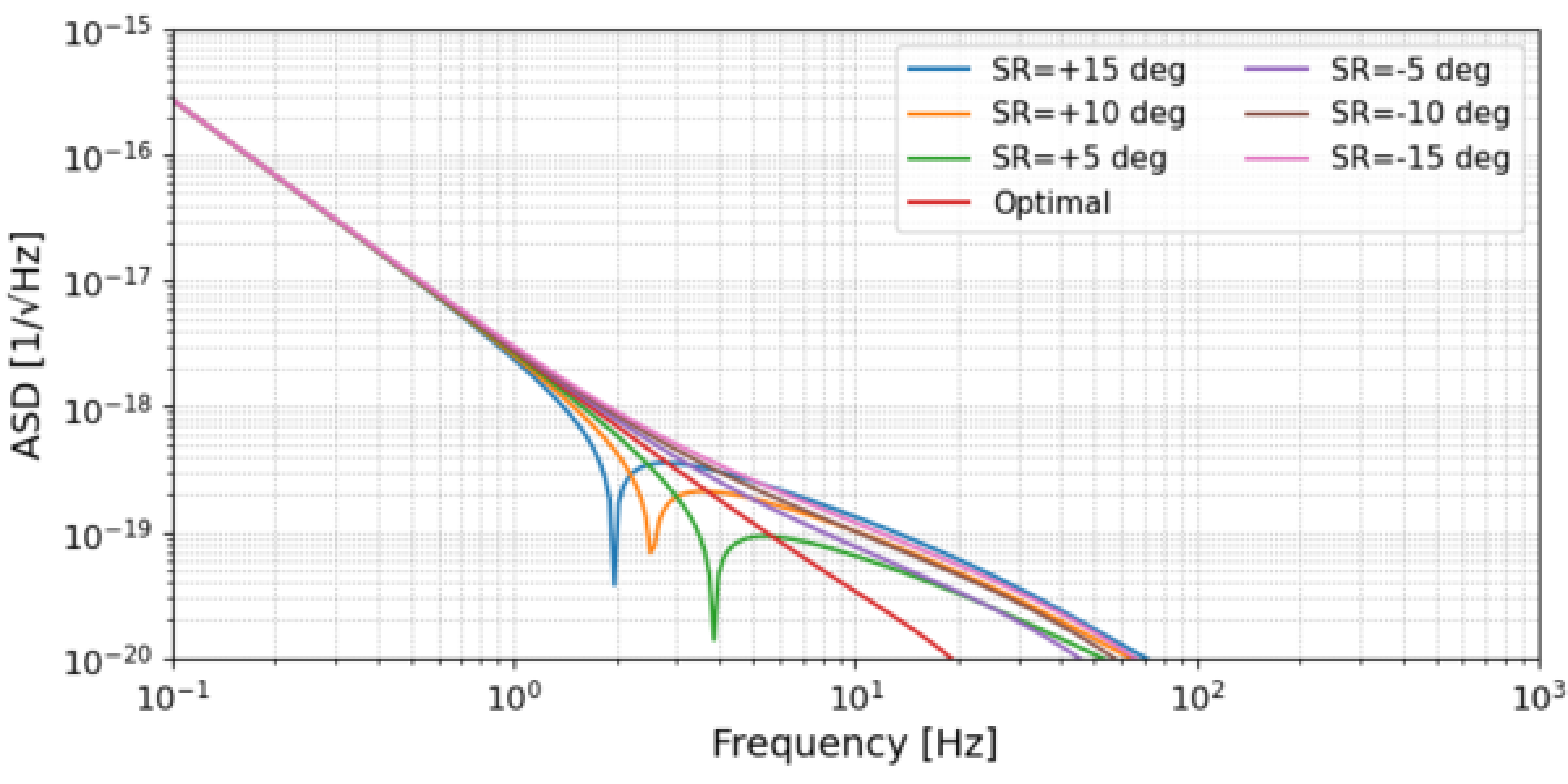}
  \caption{
  Impact of signal-recycling mirror detuning on radiation-pressure noise.  
  The turnover frequency from $1/f^2$ to $1/f$ scaling shifts depending on the detuning angle.
    The optimal detuning angle is 0 degree in this plot.
  }
  \label{fig:SR_ASD_R}
\end{figure}

The corresponding radiation-pressure-noise spectra are shown in Fig.~\ref{fig:SR_ASD_R}.
In the Sagnac speed-meter topology,
low-frequency radiation-pressure noise is suppressed through quantum correlations between measurement noise and back-action noise.
SRC detuning modifies these correlations and weakens the cancellation mechanism.
Consequently, the radiation-pressure-noise level increases at low frequencies,
leading to a degradation of the sub-hertz sensitivity that is central to the CHRONOS science goals.

\begin{figure}[t]
  \centering
  \includegraphics[width=0.8\textwidth]{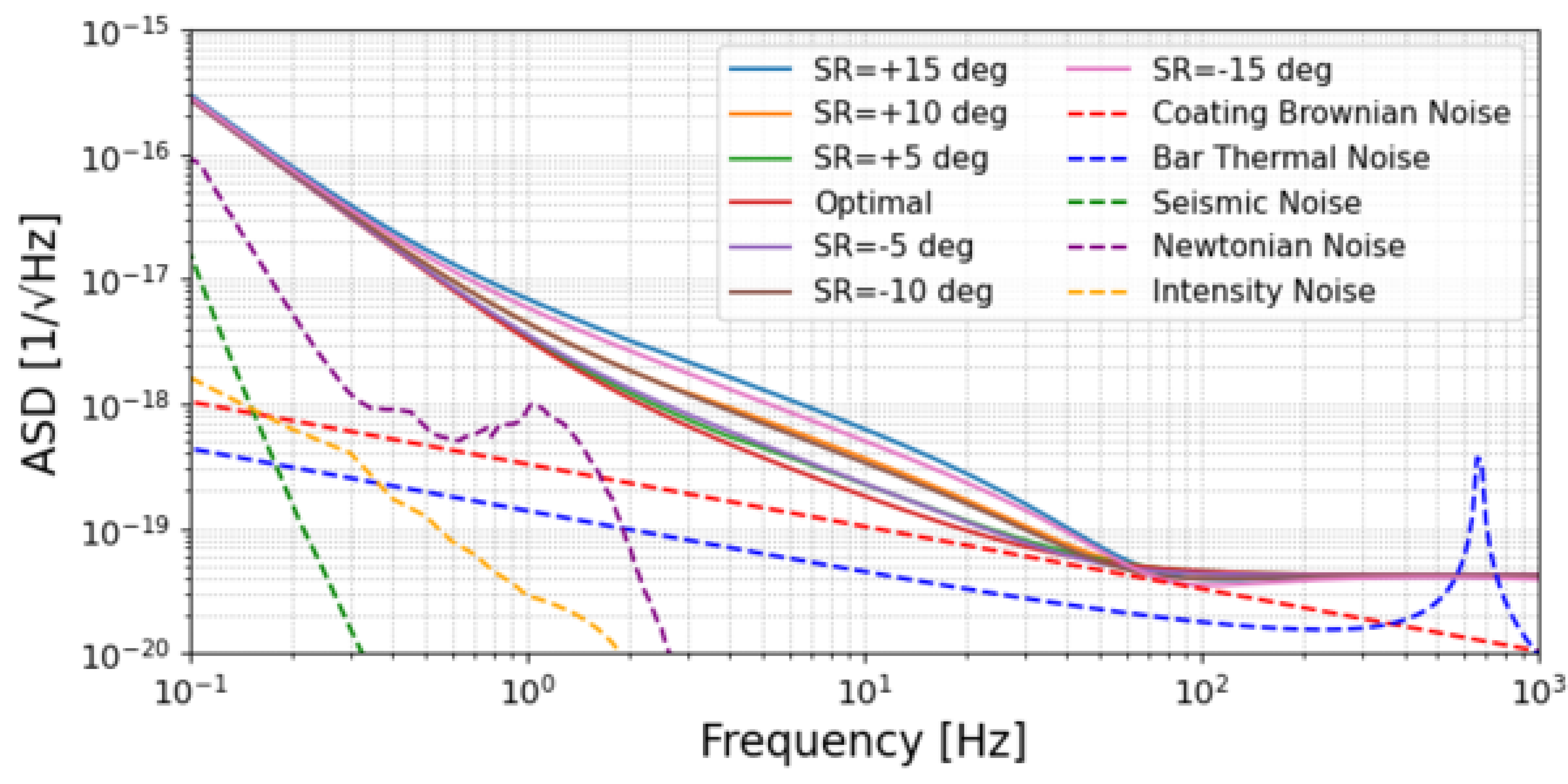}
  \caption{
  Total quantum-noise spectrum including shot noise, radiation-pressure noise, and technical noise.  
  At resonance, the three contributions balance near 1--10~Hz, while detuning leads to significant degradation.  
  The intensity noise is calculated elsewhere~\cite{Tanabe_2026}.
    The optimal detuning angle is 0 degree in this plot.
  }
  \label{fig:SR_ASD_TOT}
\end{figure}

Figure~\ref{fig:SR_ASD_TOT}
presents the resulting total sensitivity including shot noise,
radiation-pressure noise,
and other technical noise sources.
The best sensitivity is obtained at the resonant operating point,
where the speed-meter back-action suppression remains fully effective.
When the signal-recycling cavity is detuned,
the loss of quantum correlations causes a noticeable degradation in the 0.1--10~Hz band,
which dominates the overall sensitivity budget.

These results demonstrate a key difference between CHRONOS and conventional Michelson interferometers.
Whereas signal-recycling detuning is often used as an optimization parameter in position-meter detectors~\cite{BuonannoChen2001},
the speed-meter topology favors operation close to the resonant condition.
Accordingly, CHRONOS adopts the resonant SRC configuration
($\phi_s=90^\circ$ in our convention),
which preserves the quantum non-demolition response and maximizes sub-hertz sensitivity.
The signal-recycling cavity detuning angle quoted throughout this paper is therefore defined as the deviation from this $90^\circ$ reference point.

\subsection{Impact of Homodyne Detection Angle}
\label{subsec:homodyne_angle}

The homodyne detection angle $\zeta$ determines which quadrature of the output optical field is measured by the balanced homodyne detector.
By varying $\zeta$, the interferometer changes the relative weighting of the amplitude and phase quadratures,
thereby modifying the balance between shot noise and radiation-pressure noise~\cite{BuonannoChen2001,Danilishin2019}.

\begin{figure}[t]
  \centering
  \includegraphics[width=0.8\textwidth]{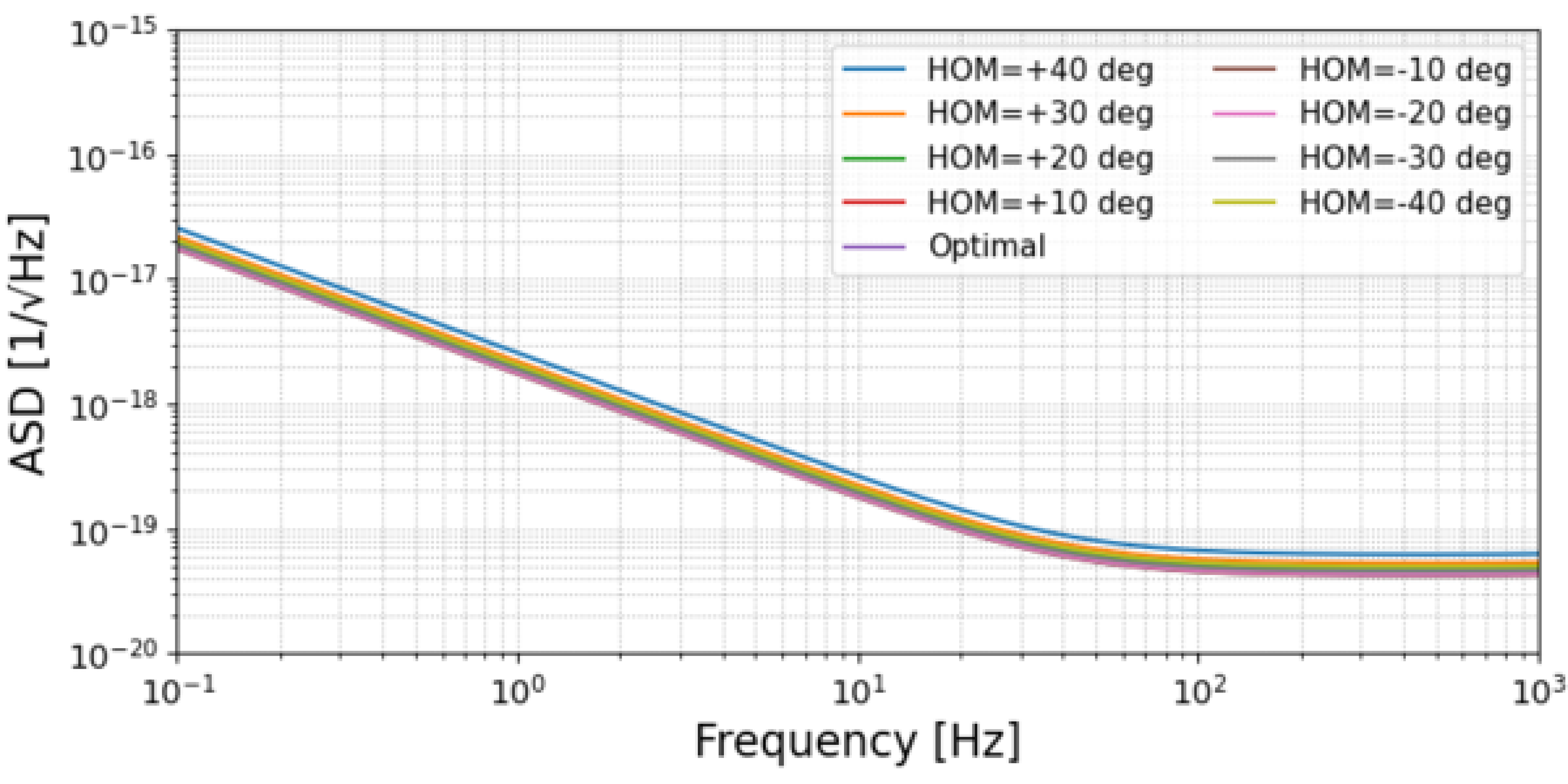}
  \caption{
 Dependence of shot noise on the homodyne detection angle $\zeta$.
Changing the homodyne angle modifies the quadrature projection of the optical field, resulting in a modest variation of the shot-noise level over the observation band.
The optimal homodyne angle is $\zeta = 46^\circ$.}
  \label{fig:HOM_ASD_S}
\end{figure}

\begin{figure}[t]
  \centering
  \includegraphics[width=0.8\textwidth]{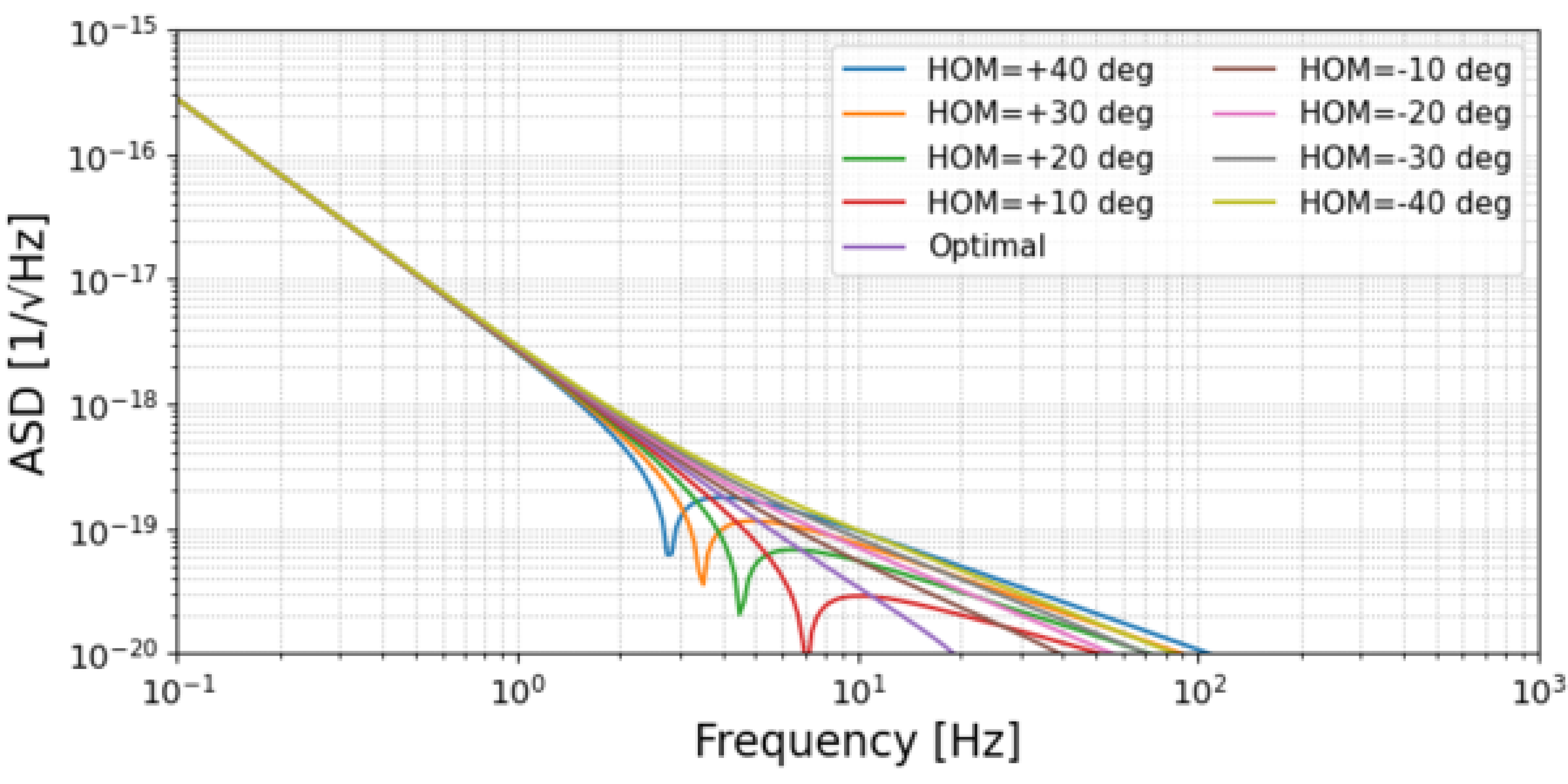}
  \caption{
Radiation-pressure noise as a function of the homodyne angle.
Amplitude--phase quadrature mixing produces interference dips in the
radiation-pressure noise spectrum.
The position and depth of these dips depend strongly on the homodyne
angle.
The low-frequency suppression observed near $\zeta = 90^\circ$
gradually weakens as the readout angle approaches the amplitude
quadrature.The optimal homodyne angle is $\zeta = 46^\circ$.}
  \label{fig:HOM_ASD_R}
\end{figure}

\begin{figure}[t]
  \centering
  \includegraphics[width=0.8\textwidth]{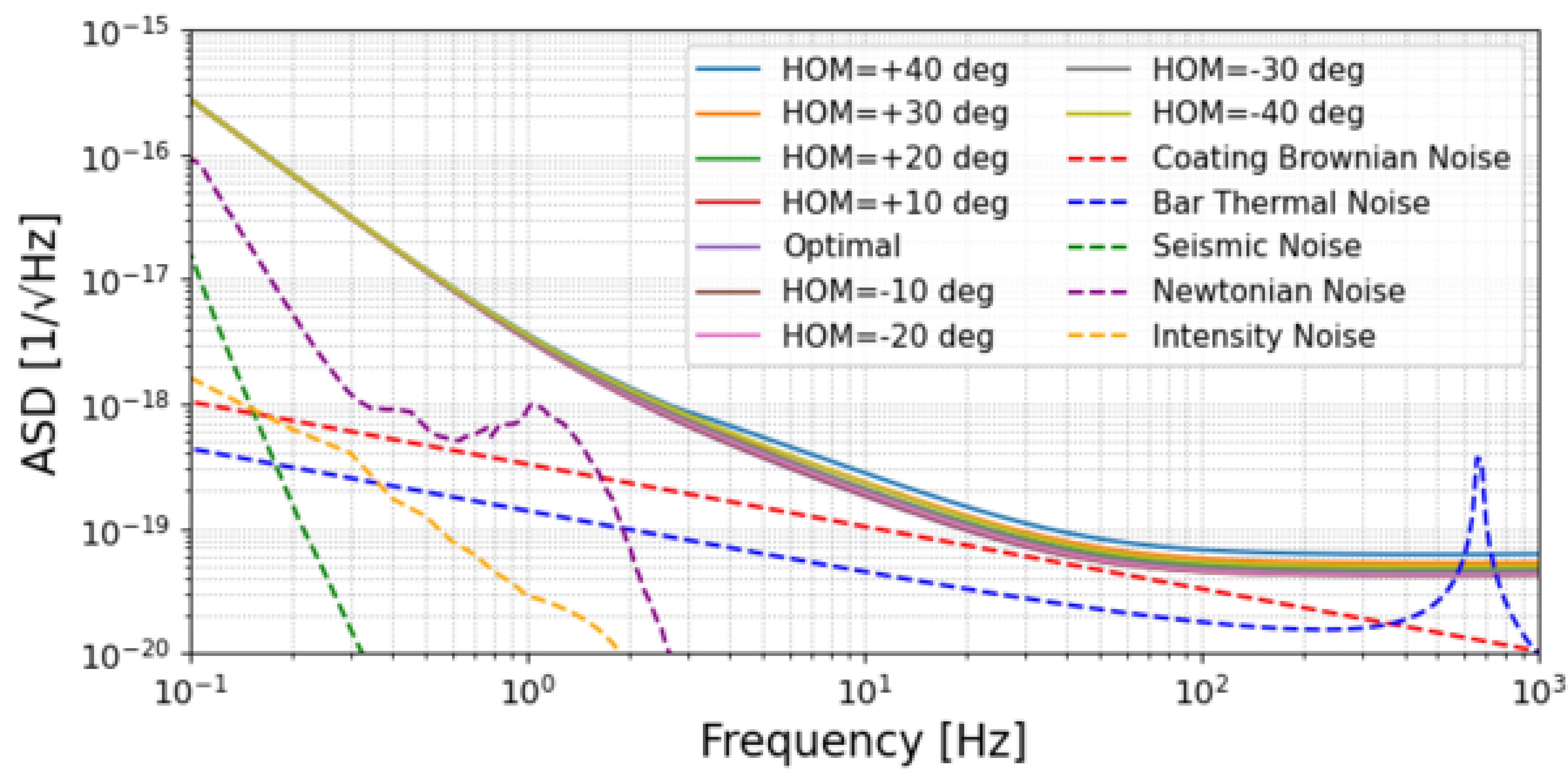}
  \caption{
  Total quantum-noise spectra for various homodyne angles.  
  The optimal configuration at $\zeta \simeq 46^{\circ}$ achieves balanced suppression of radiation-pressure and shot noise, yielding the best sensitivity near 1~Hz.  
  The intensity noise is calculated elsewhere~\cite{Tanabe_2026}. The optimal homodyne angle is $\zeta = 46^\circ$.
  }
  \label{fig:HOM_ASD_TOT}
\end{figure}

In the CHRONOS convention,
$\zeta=90^\circ$ corresponds to the phase quadrature and is adopted as the reference point for all simulations presented in this section.
The spectra shown in Figs.~\ref{fig:HOM_ASD_S}--\ref{fig:HOM_ASD_TOT}
are therefore plotted with respect to this reference angle.

Figure~\ref{fig:HOM_ASD_S}
shows the dependence of the shot-noise spectrum on the homodyne angle.
Near the phase quadrature,
the shot-noise spectrum remains nearly flat across the observation band.
Changing $\zeta$ mixes the amplitude and phase quadratures in the detected signal,
leading primarily to a uniform change in the overall shot-noise level rather than a significant modification of its frequency dependence.
The corresponding radiation-pressure-noise spectra are shown in Fig.~\ref{fig:HOM_ASD_R}.
coating-induced time-delay effects partially break the QND condition,
leading to a residual $1/f^2$ contribution.
The frequency at which this transition occurs depends sensitively on the homodyne angle.
Consequently, appropriate tuning of $\zeta$ can extend the frequency range over which back-action suppression remains effective.

The resulting total sensitivity is shown in Fig.~\ref{fig:HOM_ASD_TOT}.
An optimal homodyne angle of approximately
$\zeta\simeq46^\circ$
is found for the CHRONOS configuration.
At this angle,
shot noise and radiation-pressure noise are optimally balanced in the target observation band around $0.1$--$1~\mathrm{Hz}$,
yielding the best overall sensitivity.

The interference minima visible in the radiation-pressure-noise spectra do not appear in the total sensitivity because they are masked by the dominant shot-noise contribution.
Homodyne angles significantly smaller than $40^\circ$ or larger than $60^\circ$
lead to a noticeable degradation of sensitivity,
demonstrating that accurate control of the homodyne phase is essential for near-optimal operation of CHRONOS.

\section{Science}

\begin{figure}[t]
  \centering
  \includegraphics[width=0.7\textwidth]{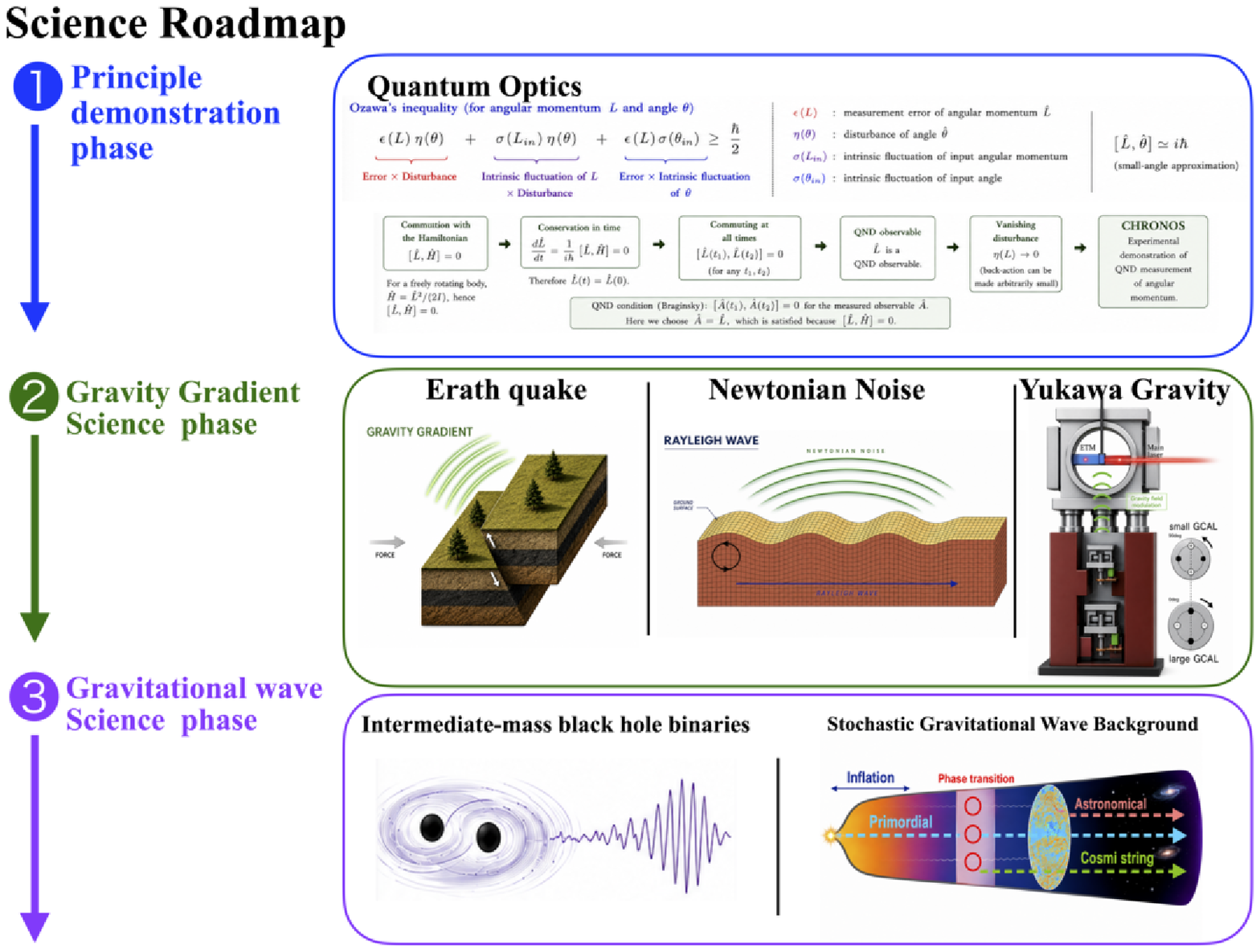}
  \caption{
Science roadmap of CHRONOS. The project is organized into three successive phases. In the principle-demonstration phase, CHRONOS aims to experimentally investigate QND measurements and test  the uncertainty relation using angular momentum and angular displacement observables. In the gravity-gradient science phase, the detector will be used to study geophysical and gravitational phenomena, including earthquake-induced gravity-gradient signals, Newtonian noise, and searches for deviations from Newtonian gravity such as Yukawa-type interactions. In the final gravitational-wave science phase, CHRONOS will target astrophysical and cosmological sources in the sub-hertz band, including intermediate-mass black-hole binaries and stochastic gravitational-wave backgrounds of primordial and astrophysical origin.
  }
  \label{fig:Roadmap}
\end{figure}

As shown in Fig.~\ref{fig:Roadmap}, the CHRONOS science program is organized into three phases, each corresponding to a different stage of detector development and scientific capability. In this section, we discuss the scientific objectives and expected sensitivities for each demonstration phase.

\subsection{Development Phase: Demonstration of Quantum Non-Demolition Measurement}
A primary objective of the initial development phase is the experimental demonstration of a QND measurement of angular momentum.
QND measurements of linear momentum have been demonstrated in recent years; however, a QND measurement of angular momentum has not yet been realized experimentally~\cite{5zyh-lg9b}. CHRONOS aims to provide the first demonstration of angular momentum system in macroscopic scale.

The characterization of this QND behavior requires a careful evaluation of both (i) the sensing function and (ii) the associated noise transfer functions, as the detector response differs from that of a conventional position meter. Calibration of low-frequency noise requires a high signal-to-noise-ratio calibration technique.
A gravity field calibrator which satisfies this requirement and provides an absolute displacement calibration, was presented in Inoue \textit{et al.}~\cite{Inoue:2026GCal}.
The resulting noise spectrum exhibits a characteristic frequency dependence, with a distinct $1/f$-like structure appearing in the 1--10~Hz band, as shown in Fig.~\ref{fig:HOM_ASD_TOT}. Observation of this signature will provide direct experimental evidence of the speed-meter response and its quantum-noise properties.

\subsection{Gravitational Field Observation Phase: Environmental and Artificial Gravitational Fields}
In the second phase, the primary science target is the observation of gravitational-field fluctuations. Detector upgrades during this stage are designed to improve the sensitivity required for environmental and artificial gravitational-field measurements.

One of the key tools in this phase is the Gravity Field Calibrator, an artificial source of a time-varying gravitational field. By utilizing Gravity Field Calibrator, signals, the calibration accuracy of the detector can be significantly improved. The principle of this technique and the expected calibration performance are summarized in Inoue et al.~\cite{Inoue:2026GCal,inoue2026probingyukawagravitymodulated, Inoue2018_GCal}.

Beyond detector calibration, the Gravity Field Calibrator, system can also be employed to search for deviations from Newtonian gravity. After coherently canceling the Newtonian gravitational component, the residual signal is entirely attributed to a possible Yukawa-type interaction. The resulting residual torque is expressed as

\begin{equation}
\tau_Y(\lambda)
=
\eta_{\rm GCAL}
GM\alpha_Y
\sum_{n=1}^{\infty}
\mathcal{Y}_n(\lambda,a,b_L,\rho,h_S,\gamma)
\sin(2\omega t),
\end{equation}

where $G$ is the gravitational constant, $M$ is the source mass, $\alpha_Y$ is the Yukawa coupling strength, $\lambda$ is the interaction range, $a$ is the source radius, $b_L$ is the baseline distance between the source and the detector, $\rho$ is the source density, $h_S$ is the source height, $\gamma$ is the geometrical opening angle of the source configuration, and $\mathcal{Y}_n$ represents the $n$-th harmonic contribution to the Yukawa interaction.

The corresponding strain-equivalent signal is then obtained through the calibrated detector response,

\begin{equation}
h_{\rm Yukawa}(\Omega;\alpha_Y,\lambda)
\simeq
\frac{2\eta_{\rm GCAL}
|\tau_Y(\Omega;\alpha_Y,\lambda)|}
{\eta_g |F_{\rm eff}| I\Omega^2},
\end{equation}

where $\tau_Y(\Omega;\alpha_Y,\lambda)$ is the Fourier component of the Yukawa-induced torque at the modulation frequency of 0.5 Hz, $\eta_g$ is the geometrical conversion factor from the torsional angle to the equivalent gravitational-wave strain, $F_{\rm eff}$ is the effective actuator response of the torsion bar, $I$ is the moment of inertia of the torsion bar, and $\Omega$ is the angular frequency. This expression converts the calculated Yukawa torque into the equivalent gravitational-wave strain using the same transfer function as employed for the standard GCal calibration. The derivation of these expressions and the detailed evaluation of the transfer functions are presented in Inoue et al.~\cite{inoue2026probingyukawagravitymodulated}.

The projected upper limit on the Yukawa coupling is then evaluated by comparing the equivalent strain with the detector sensitivity,

\begin{equation}
\alpha_{Y,\rm lim}(\lambda)
=
\frac{2\sqrt{S_h(\Omega)}}
{h_{\rm Yukawa}(\Omega;1,\lambda)\sqrt{T_{\rm eq}}},
\end{equation}

where $S_h(\Omega)$ is the detector strain noise spectrum and $T_{\rm eq}$ is the equivalent integration time determined by the residual Newtonian systematic uncertainty. Rather than assuming an arbitrary observation time, the integration is terminated when the accumulated statistical sensitivity reaches the systematic uncertainty floor.

\begin{figure}[t]
  \centering
  \includegraphics[width=0.8\textwidth]{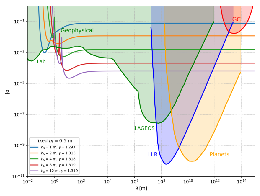}
\caption{
Projected $95\%$ confidence upper limits on the Yukawa strength parameter
$|\alpha|$ for different values of the short-baseline distance $h_S$
with the source radius fixed at $b_L=0.5~\mathrm{m}$.
For each configuration, the long-baseline distance is determined by
$h_L=\gamma h_S$, where the corresponding value of $\gamma$ is listed in
the legend.
Shaded regions indicate existing constraints from previous experiments.
}
  \label{fig:Yukawa}
\end{figure}

Figure~\ref{fig:Yukawa} presents the resulting projected constraints on Yukawa gravity for several source heights. The colored curves represent the expected CHRONOS sensitivity, while the shaded regions indicate parameter space already excluded by existing laboratory, geophysical, satellite, lunar laser ranging, planetary, and gravity calibration experiments.~\cite{Lee2017_GC,Adelberger2009,Konopliv2011,Hees2017,Borka2013,Psaltis2016,Boehle2016,Gillessen2017}. Increasing the source height enhances the gravitational modulation and improves the sensitivity over a broad range of interaction lengths. For the largest source height considered ($h_S=12~\mathrm{m}$), the projected constraint reaches a minimum coupling strength of $|\alpha|\simeq2.2\times10^{-5}$ at an interaction length of $\lambda\simeq13~\mathrm{m}$, corresponding to an equivalent strain sensitivity of $\sigma_{h_N}\simeq1.8\times10^{-20}$. Overall, CHRONOS is expected to probe Yukawa interactions over the interaction-length range from approximately $2\times10^{-1}$ to $1.3\times10^{1}~\mathrm{m}$, improving the projected sensitivity as the source height increases. The complete analysis framework and projected sensitivities are described in detail in Inoue et al.~\cite{inoue2026probingyukawagravitymodulated}.

\begin{figure}[t]
  \centering
  \includegraphics[width=0.8\textwidth]{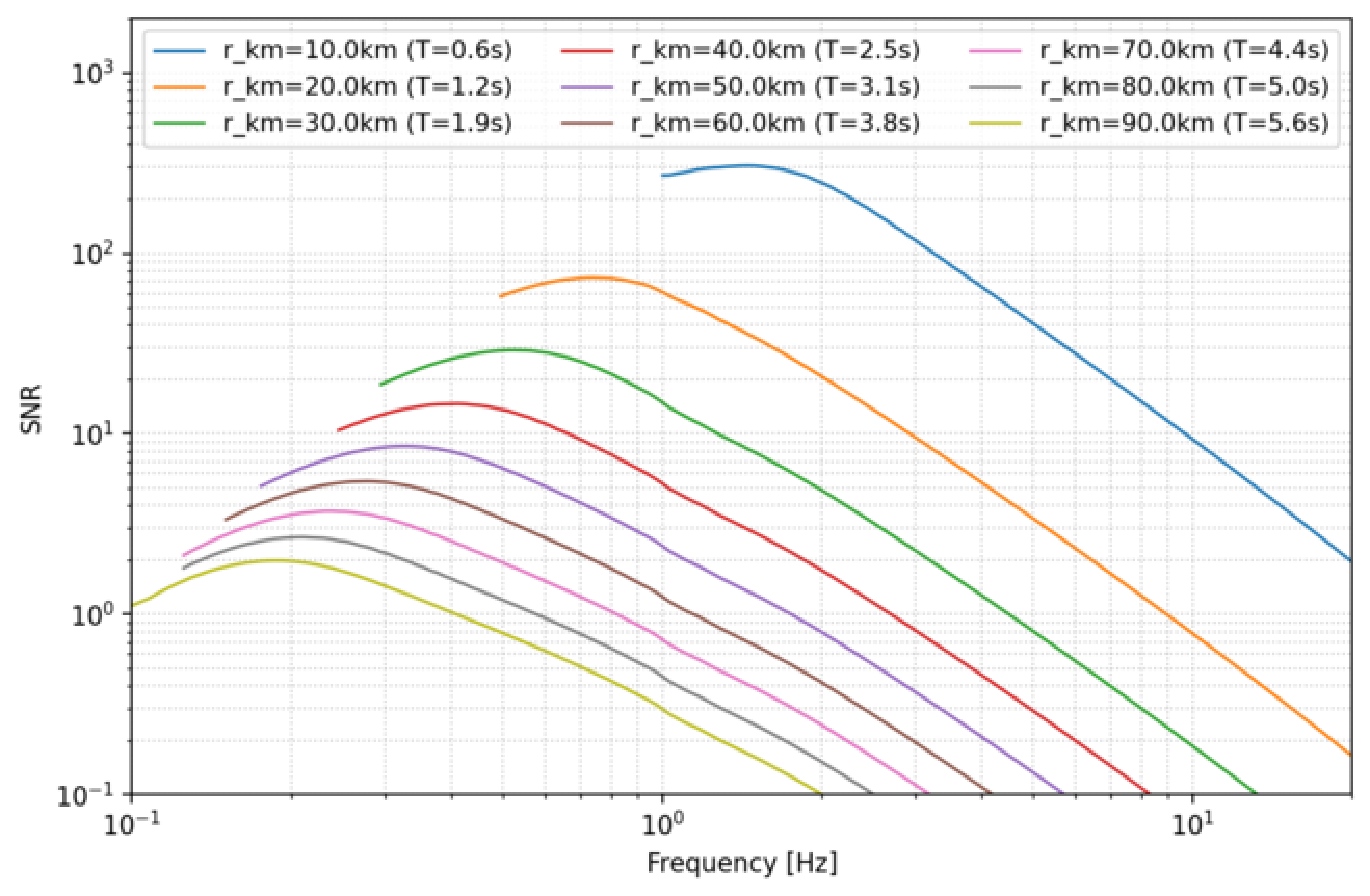}
  \caption{
Signal-to-noise ratio of earthquake-induced gravity-gradient signals as a function of frequency for different source-to-detector distances. Each curve corresponds to an earthquake occurring at a distance $r$ from the detector, with the associated seismic-wave travel time $T$ indicated in the legend. The signal-to-noise ratio is evaluated using the CHRONOS detector sensitivity and includes the frequency dependence of both the gravity-gradient signal and detector noise. Nearby events produce large signal-to-noise ratio values over a broad frequency range, while the signal amplitude decreases approximately with distance, resulting in progressively lower detectability for more distant earthquakes. The peak sensitivity is achieved in the sub-Hz to a few-Hz band, demonstrating the potential of CHRONOS for early detection and characterization of earthquake-induced gravity perturbations prior to the arrival of seismic waves.
  }
  \label{fig:Earthquake}
\end{figure}

The geographical environment of Taiwan also provides unique opportunities for
geophysical observations. Owing to the high level of seismic activity in the
region, CHRONOS can be used as a sensitive monitor of earthquakes and their
associated gravitational perturbations. Figure~\ref{fig:Earthquake} shows the
expected signal-to-noise ratio for an earthquake of magnitude M5.2 observed by
a detector located in Taipei.

The orientation-averaged prompt gravity signal induced by an earthquake is
given by
\begin{equation}
\langle h_{+}(r,t)\rangle
=
\langle h_{\times}(r,t)\rangle
=
\frac{6\sqrt{14/5}\,G}{r^{5}}
I_{4}[M_{0}](t),
\label{eq:promptgravity}
\end{equation}
where $r$ is the source-to-detector distance, $G$ is the gravitational constant,
and $I_{4}[M_{0}](t)$ denotes the fourth time integral of the seismic moment
time function.

The corresponding strain-equivalent Fourier-domain amplitude is defined as
\begin{equation}
h_{\rm EQ}(f)
=
\mathcal{F}
\left[
\langle h_{+}(r,t)\rangle
\right],
\label{eq:heq}
\end{equation}
where $\mathcal{F}[\cdot]$ denotes the Fourier transform. The detailed
calculation is summarized in Onglao et al.~\cite{onglao2026prospectsobservinggravitygradientnoise}.

The result indicates that nearby seismic events can produce detectable
gravity-gradient signals in the CHRONOS band. For source distances of
$r=10$--$40~{\rm km}$, the expected SNR exceeds 10, while events at
$r\simeq 50$--$70~{\rm km}$ remain marginally detectable with SNRs of order
a few. Even at $r\simeq 90~{\rm km}$, the signal approaches the detection
threshold. The peak sensitivity occurs around $0.3$--$1~{\rm Hz}$, where the
earthquake-induced gravity perturbation overlaps with the most sensitive
frequency region of CHRONOS.

The time delay indicated in the legend corresponds to the arrival time of the
seismic wave. Since the prompt gravity signal propagates at the speed of light,
CHRONOS could in principle observe the gravitational perturbation before the
associated seismic motion reaches the detector. For the distances shown in
Fig.~\ref{fig:Earthquake}, this lead time is approximately
$0.6$--$5.6~{\rm s}$. These results indicate that CHRONOS has the potential to
complement conventional seismometers by observing prompt gravity-gradient
signals preceding the arrival of seismic waves.

In addition, CHRONOS is expected to directly probe Newtonian noise generated by density fluctuations of Rayleigh wave. The projected sensitivity and theoretical framework are presented in Fig.~\ref{fig:HOM_ASD_TOT}. Detailed studies of the signal model, detection prospects, and expected event rates have been reported by Onglao et al~\cite{onglao2026prospectsobservinggravitygradientnoise}.

\subsection{Gravitational-Wave Observation Phase:
Intermediate-Mass Black Holes and the Stochastic Gravitational-Wave Background}

The final phase of the CHRONOS science program is dedicated to gravitational-wave astronomy in the sub-Hz band. At this stage, the detector is expected to achieve sufficient sensitivity to explore both compact-binary sources and stochastic gravitational-wave backgrounds that are inaccessible to existing observatories.

A major astrophysical target is the detection of intermediate-mass black hole binaries with component masses between stellar-mass and supermassive black holes. Although numerous black holes have been identified in both mass ranges, observational evidence for the intermediate-mass population remains limited. The sub-Hz frequency band is particularly important because it captures the inspiral phase of binaries with total masses of approximately $10^2$--$10^5\,M_\odot$, as shown in Fig.~\ref{fig:IMBH_2p5}. These systems evolve slowly compared to stellar-mass binaries observed by ground-based detectors, allowing long-duration observations and precise measurements of their orbital evolution.

The signal-to-noise ratio (SNR) is evaluated using the standard matched-filter formalism,
\begin{equation}
\mathrm{SNR}=
\sqrt{4
\int_{f_{\rm min}}^{f_{\rm max}}
\frac{|\tilde{h}(f)|^2}
{S_h(f)}
df},
\label{eq:snr}
\end{equation}
where $\tilde{h}(f)$ is the frequency-domain gravitational-wave waveform and $S_h(f)$ is the one-sided noise power spectral density of the detector. The waveform model and the calculation procedure are described in detail in Ref.~\cite{SPP-2026-3A-06,White_paper_CHRONOS}.

Our sensitivity study indicates that CHRONOS is most sensitive to intermediate-mass black hole binaries with total masses around $9\times10^3,M_\odot$. The maximum detectable luminosity distance reaches approximately $271\,\mathrm{Mpc}$ for an SNR threshold of 1. More conservatively, the corresponding horizon distances are about $90\,\mathrm{Mpc}$ for SNR = 3 and $55\,\mathrm{Mpc}$ for SNR = 5. These results demonstrate that CHRONOS can probe a substantial cosmological volume for the most favorable intermediate-mass black hole systems, providing a unique opportunity to search for a population that remains largely unexplored.

Such observations will provide valuable information on the formation channels of intermediate-mass black holes, the growth of supermassive black holes, and the assembly history of galaxies.

\begin{figure}[t]
  \centering
  \includegraphics[width=0.8\textwidth]{SNR2p5.eps}
  \caption{
Detection horizon for intermediate-mass black-hole binaries observed with the CHRONOS configuration. The luminosity distance $D_L$ is shown as a function of the total binary mass $M_{\rm tot}$. The different curves correspond to different signal-to-noise ratio thresholds used for detection. The sensitivity reaches its maximum for binaries with total masses around $10^4\,M_\odot$, where the merger and ringdown frequencies fall within the most sensitive region of the detector band. At both lower and higher masses, the observable distance decreases as the gravitational-wave signal shifts away from the optimal frequency range.}
  \label{fig:IMBH_2p5}
\end{figure}

In addition to individual compact binaries, CHRONOS will search for a stochastic gravitational-wave background. The stochastic gravitational-wave background represents a persistent gravitational-wave signal produced by the superposition of numerous unresolved sources throughout cosmic history. Astrophysical populations such as compact binaries contribute to this background, while cosmological processes in the early Universe may generate additional components. Examples include first-order phase transitions, topological defects such as cosmic strings, and other phenomena associated with physics beyond the Standard Model.

\begin{figure}[t]
  \centering
  \includegraphics[width=0.8\textwidth]{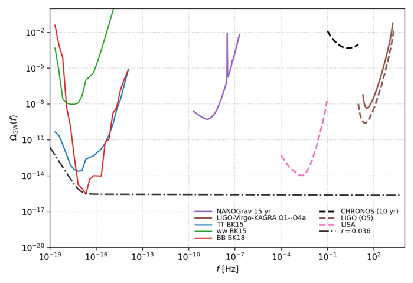}
\caption{
Comparison of the current observational constraints and projected sensitivities to the stochastic gravitational-wave background (SGWB) over a wide frequency range.
The shaded and solid curves show constraints from BK15, BK18, and NANOGrav 15 yr, while the dashed and solid sensitivity curves correspond to LISA, Advanced LIGO-Virgo-KAGRA (O5), and CHRONOS with a 10-year observation~\cite{Namikawa2019, PhysRevD.88.124032, Campeti2021Measuring}.
The horizontal dash-dotted line represents the upper limit corresponding to the tensor-to-scalar ratio of $r=0.036$.
CHRONOS is expected to bridge the frequency gap between pulsar timing arrays and ground-based laser interferometers, providing access to the sub-Hz frequency band.
}
  \label{fig:Stochastic}
\end{figure}

Figure~\ref{fig:Stochastic} presents the projected power-integrated sensitivity of CHRONOS to the stochastic gravitational-wave energy-density spectrum, $\Omega_{\rm GW}(f)$, assuming a 10-year observation time. The power-integrated sensitivity curve is calculated following the formalism of~\cite{PhysRevD.88.124032}, using the projected quantum-noise-limited strain sensitivity of the optimized CHRONOS detector over the full observation band. With the current baseline design, CHRONOS achieves its best projected sensitivity of
$\Omega_{\rm GW} \simeq 4.7\times10^{-4}$
at a frequency of approximately
$f \simeq 2.15~\mathrm{Hz},$
demonstrating its strongest capability for probing stochastic gravitational-wave backgrounds in the upper end of the sub-Hz to few-Hz frequency range.

With long-term observations, CHRONOS is expected to achieve competitive sensitivity in the frequency band between space-based and ground-based gravitational-wave detectors. This observational window is particularly important because many cosmological models predict enhanced stochastic gravitational-wave backgrounds in the sub-Hz regime. Consequently, CHRONOS will provide an important bridge between cosmic microwave background measurements at ultra-low frequencies and terrestrial interferometers operating above a few hertz, enabling a continuous exploration of the gravitational-wave Universe across an unprecedented frequency range.

\section{Discussion}

The sensitivity study presented here is optimized for the low-frequency band,
reflecting the primary motivation of CHRONOS to probe the sub-hertz regime.
In this configuration, the interferometer improves radiation-pressure-noise performance
and demonstrates the feasibility of quantum-noise-limited operation.
At the same time, it is important to emphasize that an alternative design optimized for higher frequencies
could open a different range of scientific opportunities.
By adjusting the detuning of the recycling cavities and the homodyne detection angle,
the quantum-noise spectrum can be reshaped into a shot-noise-dominated form,
enabling studies of short-duration or high-frequency signals.

Realizing such flexibility requires auxiliary optical systems in addition to the main interferometer.
In particular, a dedicated alignment system using a green laser and wavefront sensors is necessary.
Given the relatively low transmissivity of the power-recycling mirror,
such systems are essential for maintaining mode matching and cavity stability.
These subsystems are also crucial for ensuring robust operation and reliable calibration
in the presence of environmental disturbances.
While this configuration allows detailed investigation of quantum-noise shaping and control techniques,
scaling the arm length inevitably alters the balance of noise contributions.
Among these, coating thermal noise and absorption remains a decisive factor.
The choice of coating technology—mechanical loss, thickness, and deposition method—
directly determines the mid-band sensitivity floor~\cite{Harry2002,Levin1998}.
Future advances in low-loss coating materials will therefore be indispensable
for realizing the scientific goals of CHRONOS and extending the design to larger-scale detectors.

%%%%%%%%%%%%%%%%

\section{Conclusion}
\label{sec:conclusion}

In this work, we have presented the optical design and sensitivity modeling
of the CHRONOS interferometer,
a laboratory-scale triangular Sagnac speed-meter interferometer
incorporating both power- and signal-recycling techniques.
Using a combination of ABCD-matrix analysis and \textsc{Finesse3} simulations~\cite{Brown2025Finesse},
we demonstrated that stable optical eigenmodes can be realized through
optimized mirror radii of curvature and focal placement,
leading to mode-matching efficiencies exceeding $99.5\%$
across all major optical interfaces.

With the optimized optical parameters identified in this study,
the interferometer is capable of operating at the quantum-noise limit,
achieving a strain sensitivity of
\begin{equation}
    h \simeq 1\times10^{-18}~\mathrm{Hz^{-1/2}}
    \quad \text{at } 2~\mathrm{Hz}.
\end{equation}
This performance is obtained with a ring-cavity finesse of
$\mathcal{F}\simeq3.1\times10^{4}$ and a round-trip Gouy phase of
$\psi\approx153^{\circ}$,
which together provide a stable optical configuration suitable for
low-frequency precision measurements.

Our analysis shows that the detuning of the power-recycling cavity is the
dominant parameter governing the low-frequency quantum-noise performance.
An optimized power-recycling cavity detuning of
$\phi_p=5^{\circ}$ effectively suppresses radiation-pressure noise and
determines the overall quantum-noise floor.
By contrast, the signal-recycling cavity detuning
($\phi_s=0^{\circ}$) mainly introduces an approximately uniform quadrature
rotation and does not significantly improve the detector sensitivity.

We further identify an optimal homodyne readout angle of
$\zeta\simeq46^{\circ}$,
which provides the best compromise between shot noise and
radiation-pressure noise.
The results highlight that simultaneous optimization of cavity detuning
and homodyne readout is essential for realizing
quantum-noise-limited performance in the sub-Hz frequency band.

For the end test masses, a reflectivity of
$R_{\mathrm{ETM}}=99.9999\%$
is assumed throughout this study.
Optical absorption in the mirror coatings remains one of the principal
technical challenges, particularly for cryogenic operation around 10~K.
Further developments of low-absorption coating materials such as SiN and
SiON will therefore be important for extending CHRONOS toward future
long-baseline cryogenic gravitational-wave detectors.

The present study also indicates that the optimized optical configuration
effectively suppresses radiation-pressure noise in the target frequency
band while maintaining sufficient flexibility to tailor the detector
response through cavity detuning and homodyne phase.
In addition, auxiliary optical systems, including green-laser alignment
and wavefront sensing, will play an important role in preserving high
mode-matching efficiency and long-term operational stability.

{\color{black}
Beyond demonstrating the optical feasibility of the CHRONOS interferometer,
the optimized design establishes its scientific potential in the largely
unexplored sub-Hz frequency band. The projected sensitivity enables
observations of intermediate-mass black-hole binaries with total masses of
approximately $10^{2}$--$10^{5}\,M_{\odot}$, reaching a maximum detection
distance of $\sim271\,\mathrm{Mpc}$ for binaries with total masses around
$9\times10^{3}\,M_{\odot}$ (SNR = 1). For stochastic
gravitational-wave backgrounds, the detector achieves its best
power-integrated sensitivity of
$\Omega_{\rm GW}\simeq4.7\times10^{-4}$ at approximately
$2.15~\mathrm{Hz}$ after 10 years of observation. The same sensitivity also
enables competitive tests of gravity, constraining Yukawa-type deviations
from Newtonian gravity down to coupling strengths of
$|\alpha|\sim2\times10^{-5}$ at an interaction length of
$\lambda\sim13~\mathrm{m}$.

In addition to these astrophysical and fundamental-physics applications,
CHRONOS can serve as a precision geophysical observatory by detecting
prompt gravity-gradient signals generated by nearby earthquakes.
For an M5.2 event, detectable signals are expected for source distances of
up to approximately $90~\mathrm{km}$, with the gravitational perturbation
preceding the arrival of seismic waves by approximately
$0.6$--$5.6~\mathrm{s}$.
These results demonstrate that CHRONOS is not only a laboratory-scale
platform for validating quantum-noise-limited cryogenic speed-meter
interferometry, but also a versatile observatory capable of addressing key
questions in gravitational-wave astronomy, cosmology, precision tests of
gravity, and geophysics within the sub-Hz frequency band.
}

\section*{acknowledgement}
We also thank Kin-Wang Ng and Masashi Hazumi for their academic advice during the preparation of this manuscript.
We thank Masaya Hasegawa, Takahiro Kanayama, Satoki Matsushita, Hirokazu Murakami, Mai Inoue, Ryuji Shibuya, and Chia-Ming Kuo for their support in establishing the CHRONOS team.Y.I. is supported by NSTC, CHiP, and Academia Sinica in Taiwan under Grant Nos.~114-2112-M-008-006- and AS-TP-112-M01.

\appendix

\section{ABCD formalism and mode-overlap calculation}
\label{app:mode_matching}

The cavity eigenmodes and mode-matching efficiencies presented in
Sec.~\ref{sec:mode_matching}
were evaluated using the ABCD-matrix formalism for Gaussian beams.
This appendix summarizes the equations used to determine the cavity eigenmodes
and evaluate the mode-matching efficiency.

\subsection{Cavity eigenmode calculation}

The Gaussian beam is described by the complex beam parameter

\begin{equation}
\frac{1}{q(z)}
=
\frac{1}{R(z)}
-
i\frac{\lambda}{\pi w^2(z)},
\end{equation}

where $w(z)$ is the beam radius,
$R(z)$ is the wavefront radius of curvature,
and $\lambda$ is the laser wavelength.

The Rayleigh range and Gouy phase are given by

\begin{equation}
z_R
=
\frac{\pi w_0^2}{\lambda},
\end{equation}

and

\begin{equation}
\zeta(z)
=
\arctan
\left(
\frac{z}{z_R}
\right),
\end{equation}

respectively.

Propagation through an optical system is represented by the ABCD matrix

\begin{equation}
M
=
\begin{pmatrix}
A & B \\
C & D
\end{pmatrix},
\end{equation}

which transforms the beam parameter according to

\begin{equation}
q'
=
\frac{Aq+B}{Cq+D}.
\end{equation}

For a resonant cavity,
the round-trip matrix is

\begin{equation}
M_{\rm rt}
=
\prod_i M_i .
\end{equation}

The cavity eigenmode is obtained from the self-consistency condition

\begin{equation}
q
=
\frac{Aq+B}{Cq+D},
\end{equation}

which yields the beam waist,
wavefront curvature,
and Gouy phase of the resonant mode.

\subsection{Mode-overlap calculation}

The mode-matching efficiency between the injected beam
and the cavity eigenmode is quantified by the normalized overlap integral

\begin{equation}
\eta
=
\left|
\frac{
\displaystyle
\iint
E_{\rm in}^{*}
E_{\rm cav}
\,dx\,dy
}
{
\sqrt{
\displaystyle
\iint
|E_{\rm in}|^2
dx\,dy
}
\sqrt{
\displaystyle
\iint
|E_{\rm cav}|^2
dx\,dy
}
}
\right|^2 .
\end{equation}

Perfect overlap corresponds to

\begin{equation}
\eta = 1.
\end{equation}

For two Gaussian beams characterized by the beam parameters
$q_1$ and $q_2$ on a reference plane,
the overlap efficiency can be written as

\begin{equation}
\eta
=
\left|
\frac{
2\sqrt{
\Re(1/q_1)
\Re(1/q_2)
}
}
{
1/q_1
+
1/q_2^{*}
}
\right|^2 .
\end{equation}

For coaxial Gaussian beams,
the overlap can also be expressed in terms of the beam radii
and wavefront curvatures~\cite{Bayer1984}

\begin{equation}
\eta =
\frac{4}{
\left(
\frac{w_1}{w_2}
+
\frac{w_2}{w_1}
\right)^2
+
\left[
\frac{\pi w_1 w_2}{\lambda}
\left(
\frac{1}{R_1}
-
\frac{1}{R_2}
\right)
\right]^2
}.
\label{eq:eta_exact}
\end{equation}

Equation~(\ref{eq:eta_exact})
provides an intuitive estimate of the degradation of mode-matching efficiency
caused by beam-size and wavefront-curvature mismatches,
and was used to evaluate the optical tolerances of the CHRONOS design.
%%%%%%%%%%%%

\section{Detailed optimization procedure}
\label{app:optimization}

The radii of curvature of the curved mirrors were optimized
to simultaneously satisfy the cavity stability condition,
minimize optical astigmatism,
and maximize the mode-matching efficiency between the cavity eigenmodes
and the external reference beam.

The optimization procedure consisted of the following steps.

\begin{enumerate}[label=(\roman*)]

\item
For both the sagittal and tangential planes,
the cavity eigenmode parameter $q_{\mathrm{cav}}$
was calculated using the ABCD-matrix formalism.
The corresponding beam radius $w(z)$ and wavefront curvature $R(z)$
were evaluated to quantify the astigmatism arising from non-normal
incidence on the curved mirrors.

\item
The mirror curvature and incidence angle were varied
to identify configurations that minimize the effective incidence angle.
Reducing the incidence angle suppresses astigmatism,
limits higher-order transverse-mode excitation,
and decreases sensitivity to optical aberrations.

\item
The geometric symmetry of the two counter-propagating Sagnac paths
was enforced by requiring identical CW and CCW eigenmodes.
This constraint guarantees equal resonance conditions,
beam parameters,
and circulating powers for the two directions.

\item
The external injection beam parameter $q_{\mathrm{in}}$
was propagated through the optical system,
and the mismatch with the cavity eigenmode was evaluated using

\begin{equation}
\Delta q
=
q_{\mathrm{in}}
-
q_{\mathrm{cav}}.
\end{equation}

The mode-matching efficiency was calculated from the Gaussian-mode
overlap described in Appendix~\ref{app:mode_matching},
and used as the optimization objective function.

\item
For each candidate solution,
the cavity stability condition

\begin{equation}
0 < g_1 g_2 < 1
\end{equation}

was verified.
The round-trip Gouy phase and transverse-mode spacing
were also evaluated to confirm that the cavity remained
within the stable and non-degenerate operating regime.

\end{enumerate}

The optimization was considered converged when the cavity stability
condition was satisfied,
the CW and CCW eigenmodes coincided within numerical precision,
and the mode-matching efficiency reached its maximum value.
The resulting parameters are summarized in
Sec.~\ref{sec:mode_matching}.

%%%%%%%%%%%
\section{Complete Optical Element Parameters} 
\label{app:optics}

Table~\ref{tab:alloptics} lists the full set of optical elements used in the 
CHRONOS interferometer, including their coordinates, incidence angles, 
radii of curvature, and reflectivities. 
For brevity, power transmissivity and absorption values are omitted; 
they can be obtained from the relation $T = 1 - R - A$, 
with $A \approx 0$.

\begin{table*}[h]
\centering
\caption{Full list of optical elements in the CHRONOS interferometer.
The corresponding optical components are shown in Fig.~\ref{fig:layout}.
Incident angles are given in degrees, curvature radii in meters,
and power reflectivity as $R$. Based on the lengths of the ring cavity, power-recycling cavity, and signal-recycling cavity, the optimized radii of curvature are listed.}
\label{tab:alloptics}
\footnotesize  
\setlength{\tabcolsep}{6pt} 
\begin{tabular}{lrrr}
\hline\hline
Name & Incident angle [$^\circ$] & Radius of curvature [m] & Reflectivity $R$ \\
\hline
$ETM_{LY}$  &  1.21 &  2.69  & 0.999999 \\
$ETM_{RY}$  &  1.21 &  2.69  & 0.999999 \\
$MTM_{LY}$  & 37.33 & $\infty$ & 0.999999 \\
$MTM_{RY}$  & 37.33 & $\infty$ & 0.999999 \\
$ITM_{LY}$  & 51.46 & $\infty$ & 0.9999   \\
$ITM_{RY}$  & 51.46 & $\infty$ & 0.9999   \\
$CM_{RY}$   &  7.57 & 0.141  & 0.999999 \\
$CM_{LY}$   &  7.57 & 0.141  & 0.999999 \\
$BS_Y$    &  0.00 & $\infty$ & 0.5      \\
$BSM_{LY}$  & 23.14 & $\infty$ & 0.999999 \\
$BSM_{RY}$  & 23.14 & $\infty$ & 0.999999 \\
$SRM3_{Y}$  & 26.27 & $\infty$ & 0.999999 \\
$SRM2_Y$  & 26.02 & 7.15  & 0.999999 \\
$SRM1_Y$  &  0.00 & $\infty$ & 0.5      \\
$PRM4_Y$  & 26.02 & $\infty$ & 0.999999 \\
$PRM3_Y$  &  3.77 & 7.02  & 0.999999 \\
$PRM2_Y$  & 45.00 & $\infty$ & 0.999999 \\
$PRM1_Y$  &  0.00 & $\infty$ & 0.9      \\
$OM_Y$    & 45.00 & $\infty$ & 0.999999 \\
$ETM_{LX}$  &  1.21 & 2.69   & 0.999999 \\
$ETM_{RX}$  &  1.21 & 2.69   & 0.999999 \\
$MTM_{LX}$  & 37.33 & $\infty$ & 0.999999 \\
$MTM_{RX}$  & 37.33 & $\infty$ & 0.999999 \\
$ITM_{LX}$  & 51.46 & $\infty$ & 0.9999   \\
$ITM_{RX}$  & 51.46 & $\infty$ & 0.9999   \\
$CM_{RX}$   &  7.57 & 0.141  & 0.999999 \\
$CM_{LX}$   &  7.57 & 0.141  & 0.999999 \\
$BS_X$    &  0.00 & $\infty$ & 0.5      \\
$BSM_{LX}$  & 23.14 & $\infty$ & 0.999999 \\
$BSM_{RX}$  & 23.14 & $\infty$ & 0.999999 \\
$SRM3_X$  & 26.27 & $\infty$ & 0.999999 \\
$SRM2_X$  &  3.77 & 7.15  & 0.999999 \\
$SRM1_X$  &  0.00 & $\infty$ & 0.5      \\
$PRM4_X$  & 26.27 & $\infty$ & 0.999999 \\
$PRM3_X$  &  3.77 & 7.02  & 0.999999 \\
$PRM2_X$  & 45.00 & $\infty$ & 0.999999 \\
$PRM1_X$  &  0.00 & $\infty$ & 0.9      \\
$OM_X$    & 45.00 & $\infty$ & 0.999999 \\
IBS    & 45.00 & $\infty$ & 0.999999 \\
IMM2   &  7.85 & $\infty$ & 0.999999 \\
IMM1   &  7.85 & $\infty$ & 0.999999 \\
OBS    & 45.00 & $\infty$ & 0.999999 \\
IMC1   &  2.25 & $\infty$ & 0.999999 \\
IMC2   &  2.25 & $\infty$ & 0.999999 \\
IMC3   &  2.25 & 40.00 & 0.999999 \\
IMC4   &  2.25 & 40.00 & 0.999999 \\
\hline\hline
\end{tabular}
\normalsize
\end{table*}

%%%%%%%%%%%%%
\section{Noise Model}
\label{app:noise}

This appendix summarizes the principal technical noise models included in the sensitivity analysis.  
All contributions are normalized to the detector output using the frequency-dependent sensing transfer function $H_{\mathrm{sens}}(f)$ and the arm length $L_{\mathrm{bar}}$.
The assumed parameters are listed in Table~\ref{tab:rayleigh_params}.

\subsection{Coating Brownian noise}
\label{app:coat_brownian}

Brownian motion arising from the dielectric coating of the ETMs is one of the dominant fundamental noise sources in the mid-frequency band of CHRONOS. This arises from microscopic mechanical dissipation in the high-reflectivity multilayer coating, which drives thermally induced surface displacement through the fluctuation-dissipation theorem~\cite{Levin1998,Harry2002}. In the CHRONOS design, the interferometer response is ultimately limited by this process at frequencies where seismic and suspension noise are already suppressed and radiation-pressure noise is no longer dominant.

We evaluate the coating Brownian noise using the equivalent spring model in the half-infinite substrate approximation. The one-sided amplitude spectral density of the displacement noise, referred to the differential arm readout, is written as
\begin{align}
  S_{\mathrm{coat}}(\Omega) &=
  \frac{\eta_g}{L_{\mathrm{bar}}}
  \sqrt{\frac{4 k_{\mathrm{B}} T}{\pi \Omega}}
  \sqrt{\frac{1-\sigma_{s}^{2}}{\omega_{\mathrm{ETM}} Y_{s}}}
  \sqrt{\phi_{c}^{\mathrm{eff}}}.
\end{align}
Here, the effective arm length of the torsion bar is defined as $L_{\mathrm{bar}}$. The prefactor $\eta_g/L_{\mathrm{bar}}$ converts the displacement noise of the ETM surface into an equivalent strain noise. Boltzmann's constant is denoted by $k_{\mathrm{B}}$, and the coating temperature is denoted by $T$. The scaling $\propto \sqrt{T}$ explicitly shows why cryogenic operation is essential for suppressing coating Brownian noise.
The Poisson ratio and Young's modulus of the ETM substrate are denoted by $\sigma_{s}$ and $Y_{s}$, respectively. These elastic parameters determine how efficiently internal mechanical loss in the coating is converted into surface motion.
The beam width set by the spatial profile of the sensing beam on the mirror surface is denoted by $\omega_{\mathrm{ETM}}$. The effective mechanical loss angle of the coating stack is denoted by $\phi_{c}^{\mathrm{eff}}$. This term incorporates the individual loss angles, thicknesses, and elastic moduli of the high-index and low-index layers in the dielectric mirror coating. Multilayer designs with low-loss (e.g., SiN/SiON) materials reduce $\phi_{c}^{\mathrm{eff}}$ and thus directly improve sensitivity.

The $f^{-1/2}$ scaling of $S_{\mathrm{coat}}(f)$ reflects the thermally driven random-force spectrum acting on a mechanically lossy elastic system. In CHRONOS, this contribution sets the practical noise floor in the mid-frequency range (roughly $\sim 1$--$10~\mathrm{Hz}$), where radiation-pressure noise has begun to decouple due to the speed-meter topology, while seismic and suspension noise are already strongly attenuated. Consequently, coating Brownian noise defines one of the key material-driven limits to interferometer performance and directly motivates the use of cryogenic, low-loss coating technologies.

\subsection{Bar Thermal Noise}
\label{app:bar_thermal}

The torsion-bar thermal noise model combines two principal contributions:  
(1) the broadband structural damping arising from internal material loss of the suspension and bar body,  
and (2) the narrow-band angular thermal noise peaking near the torsional resonance frequency~\cite{Saulson1990,Yamamoto2006}.  
These two effects together describe both the wideband background and the resonant thermal motion of the torsion bar.

The total one-sided amplitude spectral density of the equivalent displacement noise is expressed as
\begin{align}
  S_{\mathrm{bar,bg}}(\Omega) &=
  \sqrt{\frac{4 k_{\mathrm{B}} T}{\pi^{1/2} \Omega}}
  \sqrt{\frac{1-\sigma_{s}^{2}}{\omega_{\mathrm{ETM}} Y_{s}}}
  \sqrt{\phi_{s}^{\mathrm{eff}}}, \\[2pt]
  S_{\mathrm{bar,rot}}(\Omega) &=
  \sqrt{\frac{4 k_{\mathrm{B}} T}{I}}
  \sqrt{\frac{\eta^{2} \omega_{\mathrm{tb}} \phi_{0}}
  {(\omega_{\mathrm{tb}}^{2}-\Omega^{2})^{2} + \Omega \omega_{\mathrm{tb}} \phi_{0}}}, \\[2pt]
  S_{\mathrm{bar}}(\Omega) &=
  \frac{\eta_g}{L_{\mathrm{bar}}}
  \sqrt{S_{\mathrm{bar,bg}}^{2}(\Omega) + S_{\mathrm{bar,rot}}^{2}(\Omega)} ,
\end{align}
where the torsional resonance angular frequency is defined as $\omega_{\mathrm{tb}}$, 
$\phi_{s}^{\mathrm{eff}}$ is the effective material loss angle, $\phi_{0}$ represents the loss parameter associated with the resonance peak, 
$\eta_g$ denotes the coupling efficiency between the bar rotation and interferometer readout, and $I$ is the moment of inertia of the torsion bar.

The first term, $S_{\mathrm{bar,bg}}(\Omega)$, represents the broadband contribution arising from structural damping and internal friction of the suspension fibers or bar material.  
It exhibits a characteristic $\Omega^{-1/2}$ dependence typical of thermally driven losses.  
The second term, $S_{\mathrm{bar,rot}}$, describes the resonant enhancement near $\omega \approx \omega_{\mathrm{tb}}$, 
reflecting the thermal excitation of the torsional eigenmode.  
At frequencies well below resonance, $S_{\mathrm{bar,rot}}$ scales approximately as $\Omega^{-1}$, 
while above resonance it decreases as $\Omega^{-3}$, consistent with the mechanical response of a torsional oscillator.

The total contribution $S_{\mathrm{bar}}(\Omega)$ is converted to strain noise by dividing by the torsion-bar arm length $L_{\mathrm{bar}}$.  
In CHRONOS, this term dominates below $\sim 0.5~\mathrm{Hz}$, 
where seismic noise and radiation-pressure noise are both suppressed.  
Minimizing the effective loss angles $\phi_{s}^{\mathrm{eff}}$ and $\phi_{0}$, 
and maximizing the mechanical quality factor $Q = 1/\phi_{0}$, 
are therefore essential design goals for achieving the sub-hertz sensitivity required for CHRONOS.

\subsection{Seismic Noise}
\label{app:seismic}

Noise due to ground displacement couples into the detector output through 
the pre-isolation system and the suspension transfer functions~\cite{Saulson1984,Harms2019}.  
This coupling is especially critical in the sub-hertz band, where seismic motion dominates all other technical noise sources.  
In this study, the ground-displacement spectrum measured at the ASGRAF~\cite{ASGRAF} site is adopted as the baseline input, 
combined with the modeled attenuation provided by the pre-isolation stage and the suspension system.  

At low frequencies ($f < 1~\mathrm{Hz}$), the system is assumed to employ a pre-isolation configuration 
analogous to that of LIGO, incorporating both passive and active isolation stages, 
followed by a multi-stage pendulum suspension.  
The overall transfer function thus includes the full yaw response of the suspension.  
For the rotational degree of freedom, the effective seismic coupling is empirically found to be roughly 
$1/100$ of the translational coupling amplitude, consistent with the TOBA analysis~\cite{Ando2010}.  

The resulting one-sided amplitude spectral density of the seismic contribution
is expressed as
\begin{align}
  S_{\mathrm{seis}}(\Omega) =
  \frac{|H_{\mathrm{yaw}}(\Omega)|}{G_{\mu g} \, G_{\mathrm{pre}}}
  S_{x}^{\mathrm{ground}}(\Omega),
\end{align}
where $S_{x}^{\mathrm{ground}}(\Omega)$ denotes the ground-displacement spectrum.
In this work, $S_{x}^{\mathrm{ground}}(\Omega)$ is calculated assuming
measured underground seismic data at the KAGRA site,
as reported by Michimura \textit{et al.}~\cite{Michimura_2017}.
The transfer function $H_{\mathrm{yaw}}(\Omega)$ represents the yaw response
of the suspended torsion bar to ground motion.
The factor $G_{\mu g}$ is the conversion gain between mechanical torque
and the equivalent gravitational acceleration,
which incorporates the bar geometry and the effective lever arm.
The term $G_{\mathrm{pre}}$ describes the frequency-dependent attenuation
of the pre-isolation system,
typically providing suppression proportional to $\Omega^{-4}$ to $\Omega^{-6}$
above the microseismic peak.
For the full CHRONOS facility, a dedicated vibration-isolation platform 
and cryogenic suspension are planned to achieve enhanced attenuation in the sub-hertz region.  
Such systems will combine multiple layers of active inertial sensing, low-frequency tilt decoupling, 
and cryogenic damping control to suppress seismic noise below $10^{-18}~\mathrm{m/\sqrt{Hz}}$ 
at frequencies approaching $0.1~\mathrm{Hz}$.  
These developments are essential for extending the CHRONOS sensitivity into the deep sub-hertz band, 
where Newtonian noise and coating thermal noise become the dominant limiting factors.

\begin{table}[t]
\centering
\caption{Parameters used for the noise calcuation}
\label{tab:rayleigh_params}
\begin{tabular}{lll}
\hline
Parameter & Symbol & Value \\
\hline
Gravitational constant & $G$ &
$6.67430\times10^{-11}\,\mathrm{m^3\,kg^{-1}\,s^{-2}}$ \\
Boltzmann constant&$k_{\mathrm{B}}$&$1.380649 \times 10^{-23}\ \mathrm{J\,K^{-1}}$\\
Mean ground density & $\rho_0$ &
$2000\,\mathrm{kg/m^3}$ \\
Rayleigh coupling factor & $\gamma_{\rm R}$ &
$0.83$ \\
Rayleigh phase velocity & $c_{\rm R}$ &
$3500\,\mathrm{m/s}$ \\ 
Depth of observatory & $d$ &
$200\,\mathrm{m}$ \\
Angular average & $\langle|\cos\theta|\rangle_{\rm rms}$ &
$1/\sqrt{2}$ \\
Orientation factor & $\mathcal{O}_{\rm R}$ &
$\sqrt{3/8}$ \\
Young's modulus&$Y_s$&425GPa\\
Poisson ratio&$\sigma_s$&0.1\\
Resonant frequency of ETM stage&$\omega_{ETM}$& $2\pi \times 2.0 \times 10^{-3}$ \\
Internal mode frequency&$\omega_{tb}$& $2\pi \times 667$ \\
Pre-isolation stage gain&$G_{pre}$&1000 \\
Temperature of end test mass&$T$&8 K \\
Wire mechanical loss angle&$\phi_{0}$&$1.0\times10^{-7}$\\
Coating mechanical loss angle&$\phi_c^{eff}$&$1.0\times10^{-6}$ \\
Substrate mechanical loss angle&$\phi_s^{eff}$&$1.0\times10^{-7}$ \\
\hline
\end{tabular}
\end{table}

\bibliographystyle{iopart-num}
\bibliography{chronos_optics}
\end{document}